\documentclass[12pt,onecolumn]{IEEEtran}
\usepackage{makecell,amsbsy,amsmath,amssymb,epsfig,bbm,mathrsfs,multirow,amsthm,bm}
\usepackage{array,multirow,graphicx}
\usepackage[english]{babel}
\usepackage[justification=centering]{caption}
\usepackage{url}
\usepackage{threeparttable}
\usepackage{epstopdf}
\usepackage{subfigure}
\usepackage[ruled,linesnumbered]{algorithm2e}
\usepackage{color,xcolor}
\usepackage{hyperref}

\DeclareMathAlphabet{\mathpzc}{OT1}{pzc}{m}{it}

\usepackage{url}

\IEEEoverridecommandlockouts

\begin{document}
\bibliographystyle{IEEE2}

\title{Covert D2D Communication Underlaying Cellular Network: A System-Level Security Perspective}

\author{Shaohan Feng,~\IEEEmembership{Member,~IEEE}, Xiao Lu,~\IEEEmembership{Member,~IEEE}, Kun Zhu,~\IEEEmembership{Member,~IEEE}, Dusit Niyato,~\IEEEmembership{Fellow,~IEEE}, and Ping Wang,~\IEEEmembership{Fellow,~IEEE}}

\maketitle

\vspace{-15mm}
\begin{abstract}\vspace{-2mm}
To meet the surging wireless traffic demand, device-to-device (D2D) communication underlaying cellular networks to reuse the cellular spectrum has been envisioned as a promising solution. In this paper, we aim to secure the D2D communication of the D2D-underlaid cellular network by leveraging covert communication to hide its presence from the vigilant adversary. In particular, there are adversaries aiming to detect D2D communications based on their received signal powers. To avoid being detected, the legitimate entity, i.e., D2D-underlaid cellular network, performs power control so as to hide the presence of the D2D communication. We model the combat between the adversaries and the legitimate entity as a two-stage Stackelberg game. Therein, the adversaries are the followers and aim to minimize their detection errors at the lower stage while the legitimate entity is the leader and aims to maximize its utility constrained by the D2D communication covertness and the cellular quality of service (QoS) at the upper stage. Different from the conventional works, the study of the combat is conducted from the system-level perspective, where the scenario that a large-scale D2D-underlaid cellular network threatened by massive spatially distributed adversaries is considered and the network spatial configuration is modeled by stochastic geometry. We obtain the adversary's optimal strategy as the best response from the lower stage and also both analytically and numerically verify its optimality. Taking into account the best response from the lower stage, we design a bi-level algorithm based on the successive convex approximation (SCA) method to search for the optimal strategy of the legitimate entity, which together with the best response from the lower stage constitute the Stackelberg equilibrium. Numerical results are presented to evaluate the network performance and reveal practical insights that instead of improving the legitimate utility by strengthening the D2D link reliability, increasing D2D transmission power will degrade it due to the security concern.
\end{abstract}

\begin{IEEEkeywords}
Covert communication, D2D network, cellular network, and system-level optimization. 
\end{IEEEkeywords}


\section{Introduction}
\label{sec:introduction}

By provision direct communications among the proximate users without the involvement of base station (BS) and core network, D2D communication can reduce end-to-end (E2E) latency and broaden the opportunities for local communication, such as local media sharing~\cite{7893755}. In this context, D2D network underlaying the cellular network to reuse the cellular spectrum and offload the cellular traffic, i.e., D2D-underlaid cellular network, promises improved spectrum efficiency and enhanced network throughput. However, due to the inherent broadcast nature of the wireless medium, D2D communication is vulnerable to a series of attacks, such as eavesdropping attacks. On the other hand, the conventional security approaches are inapplicable, e.g., encryption approach~\cite{7762075}, or insufficient, e.g., information-theoretical secrecy approach~\cite{9169911}, in securing D2D communication due to the following reasons:
\begin{itemize}
\item The information-theoretic secrecy approach is to achieve a positive rate difference between the legitimate and wiretap channels and only the positive rate difference will be regarded as the secrecy rate~\cite{8723525}. This can guarantee confidentiality for only part of the transmitted data, which is insufficient from the perspective of privacy protection. 

\item It is inapplicable to implement the encryption approach in decentralized networks with random access and resource-constrained networks, such as D2D networks, due to the difficulty in the cryptographic key management and the heavy computational overhead, respectively.

\item More importantly, merely preventing data from interception does not meet the stringent requirement on data security in next-generation wireless networks, especially when revealing the occurrence of communication is crippling. For example, exposure of communication activities could bare business secrets, which may become the target of attacks.

\end{itemize}

\subsection{Covert D2D Communication}

\begin{figure}[!]
	\centering
	\includegraphics[width=1\textwidth,trim=70 200 430 130, clip]{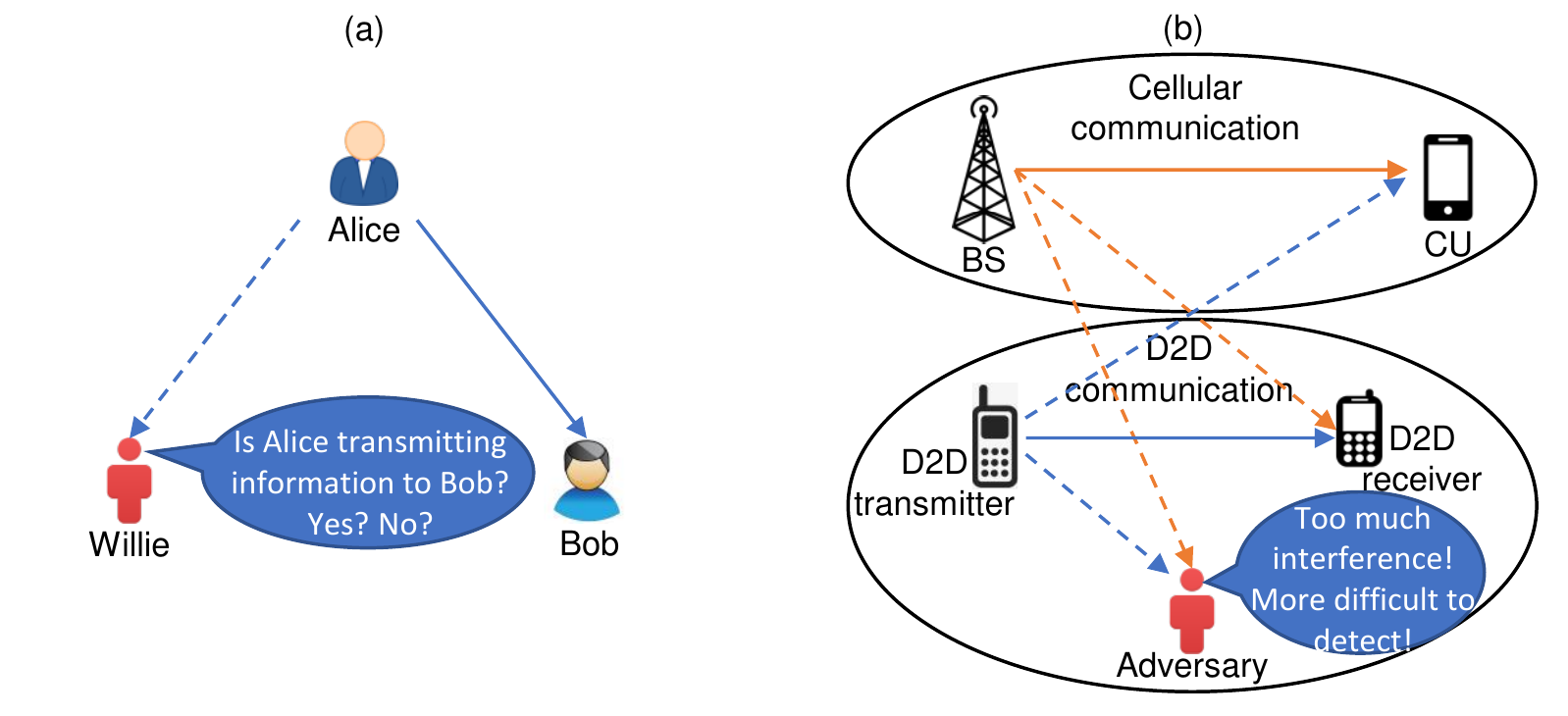}
	\caption{(a) Covert communication and (b) covert D2D communication underlaying cellular network.}
	\label{fig:covert_d2D_cellular}
\end{figure}

To cope with these challenges, we resort to the covert communication as shown in Fig.~\ref{fig:covert_d2D_cellular}(a), which aims to hide the presence of the legitimate communication between Alice and Bob from the vigilant adversary, namely Willie, while maintaining a certain legitimate communication rate. In this case, the covert communication completely secures the wirelessly transmitted data and hence outperforms the information-theoretical secrecy approach regarding the privacy protection. Moreover, different from the encryption approach, as covert communication neither relies on complicated cryptographic key management nor incurs heavy computational overhead, its feasibility in the decentralized networks and resource-constrained networks is promising. Furthermore, due to the capability that can hide the presence of the legitimate communication, the covert communication is able to avoid revealing the occurrence of the communication itself and therefore can meet the stringent requirements on data security in next-generation wireless networks. 

Motivated by the advantages of the covert communication, in this work, we adopt it to secure the D2D communication in the D2D-underlaid cellular network as shown in Fig.~\ref{fig:covert_d2D_cellular}(b). In particular, the D2D network underlays and reuses the spectrum of the cellular network for local communication, which forms a D2D-underlaid cellular network. In the meantime, the adversary aims to detect D2D communications based on its received signal power. To avoid being detected and defend against the adversary, the legitimate entity, i.e., D2D-underlaid cellular network, adopts the covert communication to hide the presence of the D2D communication. Moreover, as the signal from the cellular network, namely cellular signal, can increase the interference dynamics at the adversary, this will distort the adversary's observation and further mislead its decision-making~\cite{7805182}. In this case, the cellular signal can be leveraged to improve D2D communication covertness~\cite{feng2022securing, 9895235}. 


\subsection{System-Level Security Perspective}

As reported in~\cite{International2017}, 5G technologies will support nearly 10 million devices per ${\rm{km}}^2$ outdoors and $1000$ devices per  $100$ ${\rm{m}}^2$ indoors in numerous applications scenarios from Internet of Things (IoT) to virtual reality (VR). Such a high network densification implies that next-generation wireless networks, e.g., D2D-underlaid cellular network as one remarkable network of 5G system~\cite{7010536}, is of large-scale and requires aggressive spatial frequency reuse. On the other hand, the signal propagated over a radio link is heavily affected by location-dependent factors, such as large-scale fading, small-scale fading, and shadowing. In this case, the spatial configuration dominantly determines the system-level performance of the wireless networks. Consequently, modeling the spatial configuration of the wireless networks so as to conduct the system-level study becomes compelling. 

To address this challenge, stochastic geometry or geometric probability has been proposed. Stochastic geometry is a probabilistic analytic tool for modeling the spatial point process and hence is capable of depicting the spatial configuration of wireless networks, which has been applied to many application scenarios, such as dynamic spectrum access~\cite{7056528} and low earth orbit (LEO) satellite communication~\cite{9218989}. This depiction includes the spatial distribution of the factors that heavily affect the performance evaluation and optimization, such as the locations and transmission patterns of the network nodes, so as to reveal system-level practical insights for system development and management. In this work, we evaluate and optimize the performance of the covert communication by applying stochastic geometry to characterize the spatial configuration of the D2D-underlaid cellular network and adversaries so as to reveal system-level insights.


\subsection{Our Contributions}

In this work, we secure the D2D communication in the D2D-underlaid cellular network against the adversaries by covert communication and leveraging the existing cellular signal. We model the combat and capture the conflict of interests between the adversaries and the legitimate entity in the framework of Stackelberg game, where the legitimate entity is the leader at the upper stage and the adversaries are the followers at the lower stage. At the lower stage, the adversaries minimize their detection errors so as to accurately detect the D2D communication. At the upper stage, the legitimate entity maximizes its utility subject to the constraint on the D2D communication covertness and the cellular QoS. Therein, we include the constraint on the cellular QoS to ensure that in the D2D-underlaid cellular network, the signal from the D2D communication, which becomes a new source of interference, will not significantly degrade the performance of the cellular network.

To the authors' best knowledge, this is the first work investigating the covert communication based security solution for the D2D-underlaid cellular network from a system-level perspective. The practical insights revealed in this work, e.g., higher D2D transmission power does not imply better system performance, especially from the security perspective, can be exploited in the design of the security protocol and will contribute to the development of the security technologies. The key contributions of this work are summarized as follows:
\begin{itemize}
\item We secure D2D communication in D2D-underlaid cellular network via covert communication such that not only the locally shared data can be completely secured but also the presence of the D2D communication can be hidden. This enhances data security from the perspective of privacy protection. 

\item We characterize the spatial configuration of the D2D-underlaid cellular network by using the stochastic geometry and analytically derive the network performance metrics so as to conduct the study from the system-level perspective.

\item We model the combat and conflict of interests between the adversaries and legitimate entity, i.e., D2D-underlaid cellular network, in the framework of a two-stage Stackelberg game. Therein, the adversaries aim to minimize their detection errors at the lower stage and the legitimate entity aims to maximize its constrained utility at the upper stage.

\item We conduct equilibrium analysis and design a bi-level algorithm based on SCA to obtain the equilibrium strategies for the adversaries and legitimate entity. We also verify the optimality of the obtained equilibrium strategies. The numerical results are presented to evaluate the network performance and reveal practical insights. 
\end{itemize}

\subsection{Organization of the Paper}

The symbols used in this paper have been summarized in Table~\ref{tab:notation_value} together with their descriptions and default values. Section~\ref{sec:related_work} reviews the state-of-art related works. Section~\ref{sec:system_model} presents the system models, including the performance metric derivation. Section~\ref{sec:formulation_analysis} presents the mathematical model of the game formulation and the equilibrium analysis as well as the algorithm design. Section~\ref{sec:performance} evaluates the network performance, and Section~\ref{sec:conclusion} draws the conclusion and summarizes the insights.


\section{Related Works}
\label{sec:related_work}

\subsection{D2D Communication}

Distinctly advanced in latency, throughput, energy efficiency, and coverage extension without deploying additional infrastructure, D2D communications have attracted growing attention from the wireless research community and industry. In this case, numerous works that focus on the performance enhancement for D2D communications have been delivered~\cite{9302723}. For example, in~\cite{6953066}, two, i.e., a centralized and a distributed, power control algorithms have been developed to improve the D2D-underlaid cellular network regarding the sum rate and coverage probability, and the network spatial configuration is modeled by stochastic geometry. Regarding the centralized power control algorithm, its goal is to maximize the number of the underlaying D2D links subject to the constraint on the cellular coverage probability. In contrast, the distributed power control algorithm is to maximize the sum rate of D2D links. In~\cite{6909030}, the authors investigated two fundamental issues, i.e., spectrum access and mode selection, in the D2D enhanced cellular network. Therein, a weighted proportional fair utility function is proposed to optimize different spectrum access methods, i.e., overlay and underlay, and modes, i.e., D2D and cellular communications. In addition, the data security of D2D communication is a subject of growing concern due to the decentralized network architecture. In~\cite{7944601}, the authors developed an energy-efficient secure communication scheme based on information-theoretical secrecy approach. Different from other works that merely focus on the security perspective, a weighted product of the secrecy energy efficiency and secrecy spectral efficiency is derived and used as the performance metric. By such, an enhancement over the information secrecy, energy efficiency, and spectral efficiency can be simultaneously achieved. A resource allocation scheme is proposed in~\cite{9625847} with the purpose of maximizing the security-aware energy efficiency for simultaneous wireless information and power transfer (SWIPT)-enabled D2D users. A D2D underlaying cognitive cellular network is investigated in~\cite{7320989}, where new analytic expressions for the secrecy outage probability and secrecy throughput are derived to assess the secrecy performance in two different receiver selection schemes, i.e., selecting the receiver with either the strongest channel or nearest location. Besides, the impact of the underlaying D2D network on the system performance of the uplink cellular network has also been investigated in~\cite{6928445} by presenting a comprehensive and tractable analytical framework for D2D-enabled uplink cellular networks. Flexible mode selection scheme is adopted for the D2D communications and, moreover, truncated channel inversion power control is applied so as to ensure a certain power for the received signal at the intended receiver.


\subsection{Covert Communication}

Due to the capability that can hide the presence of wireless communication, covert communication has gained growing attention and is envisioned as a new paradigm in data security for next-generation wireless networks. To date, covert communication has been applied to secure data transmission in numerous application scenarios~\cite{9580594, 9524501, 9456902, 9662051}, and the enhanced method of which has also been widely investigated~\cite{9361424, 8654724, 9524501, 8600757}. In~\cite{9361424}, a friendly jammer is deployed to assist in achieving covert communication for legitimate communication and defending against an observant adversary. The artificial noise (AN) emitted by the friendly jammer can distort the observation of the adversary and further mislead its decision, thereby enhancing communication covertness. Multi-antenna technology has been adopted in~\cite{8654724} to achieve covert communication. Therein, the capability of the multi-antenna technology that can achieve constructive (destructive) effect for the signal in the desired (undesired) direction has been leveraged to reduce the signal leakage at the adversary and hence induce a low probability of detection. In~\cite{9524501}, a two-user non-orthogonal multiple access (NOMA) system has been investigated, where these two users in the NOMA system have been classified as one covert user and one public user, and the signal from the public user can help to hide the presence of the signal of the covert user. Similar to~\cite{9108996}, to further improve communication covertness, an intelligent reflecting surface is deployed to weaken the signal leakage at the adversary for the covert user. The interference is leveraged in~\cite{8600757} to hide the presence of the IoT transmission and enable highly sensitive information delivery, where the interference in the IoT system works similarly to the AN in~\cite{9361424} and the signal from the public user in~\cite{9524501}. An unmanned aerial vehicle (UAV)-enabled military surveillance scenario has been studied in~\cite{9456902}, where an UAV has been deployed to surveil an area of interest. In this scenario, covert communication has been applied to hide the presence of the UAV's wireless communication against from army while maintaining a high-level communication quality at the intended (legitimate) receiver. An UAV-Aided Wireless-Powered IoT system has been studied in~\cite{9662051}. An interesting trade-off between the freshness of the IoT data, i.e., requiring high transmission power, and the communication covertness of the data aggregation process, i.e., upper bound the transmission power, has been investigated. 


\subsection{Stochastic Geometry}

Wireless networks are fundamentally limited by the network geometrical configuration due to the fact that the spatial locations of network nodes strongly affect the intensity of the received signal and interference~\cite{5226957}. To tractably depict the network geometrical configuration and thereafter evaluate the corresponding communication-theoretic results of the wireless networks, stochastic geometry has been developed and widely adopted~\cite{haenggi2012stochastic}. For example, K-tier downlink heterogeneous cellular networks have been modeled and analyzed in~\cite{6171996}. Therein, the spatial configuration of each tier is modeled by using homogeneous Poisson point process (PPP) while with different cross-tier transmission powers, supported data rates, and spatial densities. Interesting insight has been revealed that the signal-to-interference-plus-noise ratio (SINR) outage probability is independent of the number of tiers and the density of BSs when all the tiers have the same target SINR. Similarly, stochastic geometry has been adopted to simplify the interference characterization in multi-tier cellular wireless networks~\cite{6524460}. Therein, tractable and accurate performance bounds for the multi-tier and cognitive cellular wireless networks are provided to reveal practical insights in the system design. Stochastic geometry is adopted to model a large-scale mobile ad-hoc network in order to optimize a variant of the spatial ALOHA protocol, i.e., opportunistic ALOHA protocol~\cite{5226963}. The advantages of the optimized variant over the original protocol have been both analytically and numerically demonstrated in terms of significant improvement in mean throughput per unit area. In~\cite{7056528}, the spatial configuration of the cellular network with the cognitive and energy harvesting based-D2D communication has been modeled and analyzed by using stochastic geometry. The enhancement in the D2D outage probability due to the cognitive channel access scheme has been validated for both the random and prioritized spectrum access policies. To evaluate the theoretical performance improvement of adopting multicell cooperation in dense cellular networks for mitigating intercell interference, stochastic geometry has been applied to model an entire plane of interfering cells in~\cite{6376184}, where the service outage probability has been derived as a function of mobile locations, scattering environment, and cooperative patterns among BSs.

Motivated by these related works, we apply covert communication to secure the D2D communication in the D2D-underlaid cellular network and model the network spatial configuration by stochastic geometry so as to conduct the study from the system-level perspective.


\section{System Model}
\label{sec:system_model}

\begin{figure}[!]
	\centering
	\includegraphics[width=1\textwidth,trim=150 100 230 140, clip]{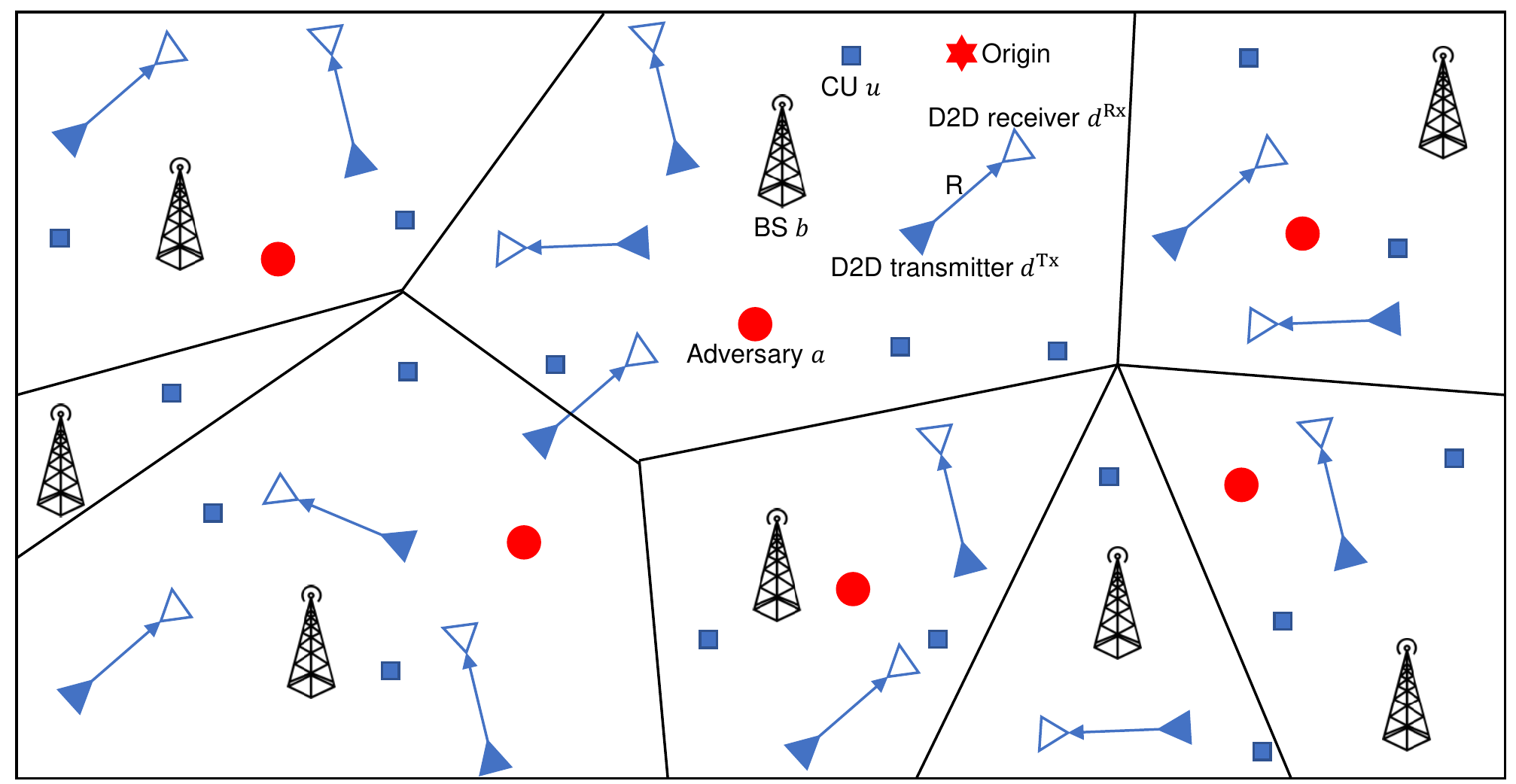}
	\caption{D2D network with D2D receivers and D2D transmitters marked by blue hollow triangles and blue solid triangles, respectively, underlaying the cellular network with CUs marked by blue rectangles and BSs threatened by adversaries marked by the red circles.}
	\label{fig:system_model}
\end{figure}

We consider a D2D-underlaid cellular network as shown in Fig.~\ref{fig:system_model}, where a D2D network underlays a downlink cellular network and operates over the same time-frequency resource block (RB). For the D2D network, its D2D transmitters follow a homogeneous PPP $\Phi_{{\cal{D}}^{\rm{Tx}}}$ of density $\lambda_{{\cal{D}}^{\rm{Tx}}}$, each having a dedicated D2D receiver located at distance $R$ in a random orientation. These D2D transmitters and their dedicated D2D receivers constitute D2D users (pairs) and form a Poisson bipolar network~\cite{haenggi2012stochastic}. For the cellular network, the BSs follow an independent homogeneous PPP $\Phi_{\cal{B}}$ of density $\lambda_{\cal{B}}$. Similar to~\cite{7893755}, we consider a nearest-BS association policy for the cellular users (CUs) and the CUs are following a homogeneous PPP $\Phi_{{\cal{U}}}$ of density $\lambda_{\cal{U}} \gg \lambda_{\cal{B}}$ such that each BS has at least one CU in its Voronoi cell. ALOHA-type channel access scheme is adopted for both D2D and cellular networks\footnote{Here, we consider that the BS follows ALOHA-type channel scheme to be active over the considered RB.}, where each D2D transmitter and BS are active independently over the considered RB with the probabilities of ${\mathbb{P}}^{{\cal{D}}_1}$ and ${\mathbb{P}}^{{\cal{C}}_1}$, respectively, and the events of which are denoted by ${\cal{D}}_1$ and ${\cal{C}}_1$, respectively~\cite{7893755}. The events that the D2D transmitter and BS are mute over the considered RB are denoted by ${\cal{D}}_0$ and ${\cal{C}}_0$, respectively. For the adversaries, their spatial distribution follows an independent PPP $\Phi_{\cal{A}}$ of density $\lambda_{\cal{A}}$~\cite{9736993}. To analyze the network performance, we consider the D2D receiver and CU nearest to the origin as the typical D2D receiver and typical CU, respectively, which are denoted by D2D receiver $d^{\rm{Rx}}$ and CU $u$, respectively. Accordingly, the D2D transmitter associated by the typical D2D receiver is the typical D2D transmitter, denoted by D2D transmitter $d^{\rm{Tx}}$. Similarly, the BS associated by the typical CU is defined as the typical BS, denoted by BS $b$. Practically, D2D transmitter $d^{\rm{Tx}}$ will defend against its nearest (most threatening) adversary, which is therefore defined as the typical adversary and denoted by adversary $a$, from transmission detection. 

\subsection{Channel Model}

For the channel, we define it between the typical D2D transmitter and typical D2D receiver, i.e., D2D transmitter $d^{\rm{Tx}}$ and D2D receiver $d^{\rm{Rx}}$ locating at ${\bf{x}}_{d^{\rm{Tx}}}$ and ${\bf{x}}_{d^{\rm{Rx}}}$, respectively, as an example and the channel between any other two network nodes can be defined similarly. The channel between D2D transmitter $d^{\rm{Tx}}$ and D2D receiver $d^{\rm{Rx}}$ is $h_{d^{\rm{Tx}} {d^{\rm{Rx}}}} \ell\left({\bf{x}}_{d^{\rm{Tx}}}, {\bf{x}}_{d^{\rm{Rx}}}\right)$, where $h_{d^{\rm{Tx}} {d^{\rm{Rx}}}}$ captures the small-scale fading and $\ell\left({\bf{x}}_{d^{\rm{Tx}}}, {\bf{x}}_{d^{\rm{Rx}}}\right) \triangleq \left\|{\bf{x}}_{d^{\rm{Tx}}} - {\bf{x}}_{d^{\rm{Rx}}}\right\|^{-\alpha} = r_{d^{\rm{Tx}} d^{\rm{Rx}}}^{-\alpha} = R^{-\alpha}$ measures the large-scale fading with $\alpha$ and $\left\|\cdot\right\|$ being the path-loss exponent and Euclidean distance operator, respectively. Rayleigh fading, i.e., independent and identical exponential distribution with a unit mean, is assumed for the small-scale fading of all the channels~\cite{haenggi2012stochastic}, e.g., $h_{d^{\rm{Tx}} {d^{\rm{Rx}}}} \sim \exp\left(1\right)$. 


\subsection{Network Model}
\label{subsec:network_model}

\subsubsection{Performance Metrics for the D2D Network}

The received signal power at the typical D2D receiver, i.e., D2D receiver ${d^{\rm{Rx}}}$, regarding the activation status of its associated D2D transmitter, i.e., D2D transmitter ${d^{\rm{Tx}}}$, is 
\begin{equation}\label{eq:signal_model_d_RX}
y_{d^{\rm{Rx}}} = \left\{
\begin{aligned}
& p^{\rm{D}} h_{d^{\rm{Tx}} d^{\rm{Rx}}} R^{-\alpha} + I_{d^{\rm{Rx}}}^{\rm{D}} + I_{d^{\rm{Rx}}}^{\rm{C}} + N_{d^{\rm{Rx}}}, \, & {\text{if}} \, {\cal{D}}_1,\\
& I_{d^{\rm{Rx}}}^{\rm{D}} + I_{d^{\rm{Rx}}}^{\rm{C}} + N_{d^{\rm{Rx}}}, \, & {\text{if}} \, {\cal{D}}_0,\\
\end{aligned}\right.
\end{equation}
where $p^{\rm{D}}$ is the D2D transmission power, $I_{d^{\rm{Rx}}}^{\rm{D}} = p^{\rm{D}} \sum\limits_{{d^{\rm{Tx}}}' \in \left\{\left.{\cal{D}}^{\rm{Tx}} \backslash \left\{d^{\rm{Tx}}\right\}\right| {d^{\rm{Rx}}}\right\}} {\mathbbm{1}}_{{d^{\rm{Tx}}}'} h_{{d^{\rm{Tx}}}' {d^{\rm{Rx}}}} \ell\left({\bf{x}}_{{d^{\rm{Tx}}}'}, {\bf{x}}_{d^{\rm{Rx}}}\right)$ and $I_{d^{\rm{Rx}}}^{\rm{C}} = p^{\rm{C}} \sum\limits_{b' \in \left\{\left.{\cal{B}} \right| {d^{\rm{Rx}}}\right\}} {\mathbbm{1}}_{b'} h_{b' {d^{\rm{Rx}}}} \ell\left({\bf{x}}_{b'}, {\bf{x}}_{d^{\rm{Rx}}}\right)$ are the sum of the signal powers from D2D transmitter ${d^{\rm{Tx}}}' \in \left\{\left.{\cal{D}}^{\rm{Tx}} \backslash \left\{d^{\rm{Tx}}\right\}\right| {d^{\rm{Rx}}}\right\}$ and that from BS $b' \in \left\{\left.{\cal{B}} \right| {d^{\rm{Rx}}}\right\}$, respectively, and $N_{d^{\rm{Rx}}}$ is the additive noise at D2D receiver $d^{\rm{Rx}}$. Therein, $p^{\rm{C}}$ is the cellular transmission power, and ${\mathbbm{1}}_{{d^{\rm{Tx}}}'}$ indicates the activation status of D2D transmitter ${d^{\rm{Tx}}}'$, where it equals $1$ if D2D transmitter ${d^{\rm{Tx}}}'$ is active and $0$ otherwise. Similarly, ${\mathbbm{1}}_{b'}$ is the activation indicator of BS $b'$. $\left\{{\bf{x}}_{{d^{\rm{Tx}}}'}\right\}_{{d^{\rm{Tx}}}' \in \left\{\left.{\cal{D}}^{\rm{Tx}} \backslash \left\{d^{\rm{Tx}}\right\}\right| {d^{\rm{Rx}}}\right\}} = \Phi_{\left\{\left.{\cal{D}}^{\rm{Tx}} \backslash \left\{d^{\rm{Tx}}\right\}\right| {d^{\rm{Rx}}}\right\}}$ is the point process (PP) of the D2D transmitters that excludes D2D transmitter $d^{\rm{Tx}}$ with D2D receiver $d^{\rm{Rx}}$ as the observation point and $\left\{{\bf{x}}_{b'}\right\}_{b' \in \left\{\left.{\cal{B}} \right| {d^{\rm{Rx}}}\right\}}= \Phi_{\left\{\left.{\cal{B}} \right| {d^{\rm{Rx}}}\right\}}$ is the PP of the BSs with D2D receiver $d^{\rm{Rx}}$ as the observation point. Consequently, the D2D link reliability can be measured in terms of the successful decoding probability at D2D receiver $d^{\rm{Rx}}$ as follows:
\begin{equation}\label{eq:Prob_success_D2D}
{\mathbb{P}}\left[\left. {\text{SINR}}_{d^{\rm{Rx}}} > \theta^{\rm{D}} \right| {\cal{D}}_1 \right]
= {\mathbb{P}}\left[ \frac{p^{\rm{D}} h_{d^{\rm{Tx}} d^{\rm{Rx}}} R^{-\alpha}} {I_{d^{\rm{Rx}}}^{\rm{D}} + I_{d^{\rm{Rx}}}^{\rm{C}} + N_{d^{\rm{Rx}}}} > \theta^{\rm{D}} \right],
\end{equation}
where $\theta^{\rm{D}}$ is the SINR threshold at D2D receiver for successfully decoding the received signal.

\subsubsection{Performance Metrics for the Cellular Network}

The received signal power at the typical CU, i.e., CU $u$, regarding the activation status of its associated BS, i.e., BS $b$, is 
\begin{equation}\label{eq:received_signal_u_with_D2D}
y_{u} = \left\{
\begin{aligned}
& p^{\rm{C}} h_{b u} \ell\left({\bf{x}}_{b}, {\bf{x}}_{u}\right) + I_{u}^{\rm{D}} + I_{u}^{\rm{C}} + N_{u}, \, & {\text{if}} \ {\cal{C}}_1,\\
& I_{u}^{\rm{D}} + I_{u}^{\rm{C}} + N_{u}, \, & {\text{if}} \, {\cal{C}}_0,
\end{aligned}\right.
\end{equation}
where $I_{u}^{\rm{D}} = p^{\rm{D}} \sum\limits_{{d^{\rm{Tx}}}' \in \left\{\left.{\cal{D}}^{\rm{Tx}} \right| u\right\}} {\mathbbm{1}}_{{d^{\rm{Tx}}}'} h_{{d^{\rm{Tx}}}' u} \ell\left({\bf{x}}_{{d^{\rm{Tx}}}'}, {\bf{x}}_{u}\right)$ and $I_{u}^{\rm{C}} = p^{\rm{C}} \sum\limits_{b' \in \left\{\left.{\cal{B}} \backslash \left\{b\right\}\right| u\right\}} {\mathbbm{1}}_{b'} h_{b' u} \ell\left({\bf{x}}_{b'}, {\bf{x}}_{u}\right)$ are the sum of the signal powers from D2D transmitter ${d^{\rm{Tx}}}' \in \left\{\left.{\cal{D}}^{\rm{Tx}} \right| u\right\}$ and that from BS $b' \in \left\{\left.{\cal{B}} \backslash \left\{b\right\}\right| u\right\}$, respectively, and $N_{u}$ is the additive noise at CU $u$. Therein, $\left\{{\bf{x}}_{{d^{\rm{Tx}}}'}\right\}_{{d^{\rm{Tx}}}' \in \left\{\left.{\cal{D}}^{\rm{Tx}} \right| u\right\}} = \Phi_{\left\{\left.{\cal{D}}^{\rm{Tx}} \right| u\right\}}$ is the PP of D2D transmitters with CU $u$ as the observation point and $\left\{{\bf{x}}_{b'}\right\}_{b' \in \left\{\left.{\cal{B}} \backslash \left\{b\right\}\right| u\right\}} = \Phi_{\left\{\left.{\cal{B}} \backslash \left\{b\right\}\right| u\right\}}$ is the PP of BSs that excludes BS $b$ with CU $u$ as the observation point. Due to the nearest-BS association policy of the CUs, no BS is nearer to CU $u$ than BS $b$. Hence, the distribution of the distance between CU $u$ and BS $b$, i.e., $\left\|{\bf{x}}_{b} - {\bf{x}}_{u}\right\|$, is as follows (see Section III-A of~\cite{6042301} for details)
\begin{equation}\label{eq:pro_dis_b_u}
f_{\left\|{\bf{x}}_{b} - {\bf{x}}_{u}\right\|} \left(r\right) = f_{r_{b u}} \left(r\right) = 2 \pi \lambda_{{\cal{B}}} r \exp\left( -\pi \lambda_{{\cal{B}}} r^2\right).
\end{equation} 
As the D2D network is underlaying the cellular network, we have to evaluate the impact of the interference from the D2D network on the performance of the cellular network. Here, we additionally evaluate the cellular link reliability without taking into account the interference from the D2D network, i.e.,~(\ref{eq:Prob_success_cellular_wo_D2D}). The received signal power at CU $u$ without taking into account the interference from the D2D network is as follows:
\begin{equation}\label{eq:received_signal_u_without_D2D}
{\widetilde{y}}_{u} = \left\{
\begin{aligned}
& p^{\rm{C}} h_{b u} \ell\left({\bf{x}}_{b}, {\bf{x}}_{u}\right) + I_{u}^{\rm{C}} + N_{u}, \, & {\text{if}} \, {\cal{C}}_1,\\
& I_{u}^{\rm{C}} + N_{u}, \, & {\text{if}} \, {\cal{C}}_0.
\end{aligned}\right.
\end{equation}
Then, similar to~(\ref{eq:Prob_success_D2D}), the cellular link reliability taking and without taking into account the interference from the D2D network are 
\begin{equation}\label{eq:Prob_success_cellular_w_D2D}
{\mathbb{P}}\left[\left. {\text{SINR}}_{u} > \theta^{\rm{C}} \right| {\cal{C}}_1 \right]
= {\mathbb{P}}\left[ \frac{p^{\rm{C}} h_{b u} \ell\left({\bf{x}}_{b}, {\bf{x}}_{u}\right)} {I_{u}^{\rm{D}} + I_{u}^{\rm{C}} + N_{u}} > \theta^{\rm{C}} \right]
\end{equation}
and
\begin{equation}\label{eq:Prob_success_cellular_wo_D2D}
{\mathbb{P}}\left[\left. {\widetilde{\text{SINR}}_{u}} > \theta^{\rm{C}} \right| {\cal{C}}_1 \right]
= {\mathbb{P}}\left[ \frac{p^{\rm{C}} h_{b u} \ell\left({\bf{x}}_{b}, {\bf{x}}_{u}\right)} { I_{u}^{\rm{C}} + N_{u}} > \theta^{\rm{C}} \right],
\end{equation}
respectively, where $\theta^{\rm{C}}$ is the SINR threshold at CU for successfully decoding the received signal.

\subsubsection{Performance Metrics for the Adversary}

The received signal power at the typical adversary, i.e., adversary $a$, regarding the activation status of its target D2D transmitter, i.e., D2D transmitter ${d^{\rm{Tx}}}$, is 
\begin{equation}\label{eq:signal_model_a}
y_{a} =
\left\{\begin{aligned}
& p^{\rm{D}} h_{{d^{\rm{Tx}}} a} \ell\left({\bf{x}}_{d^{\rm{Tx}}}, {\bf{x}}_{a}\right) + I_{a}^{\rm{D}} + I_{a}^{\rm{C}} + N_{a}, \, & {\text{if}} \, {\cal{D}}_1,\\
& I_{a}^{\rm{D}} + I_{a}^{\rm{C}} + N_{a}, \, & {\text{if}} \, {\cal{D}}_0,\\
\end{aligned}\right.
\end{equation}
where $I_{a}^{\rm{D}} = p^{\rm{D}} \sum\limits_{{d^{\rm{Tx}}}' \in \left\{\left.{\cal{D}}^{\rm{Tx}} \backslash \left\{{d^{\rm{Tx}}}\right\}\right| a\right\}} {\mathbbm{1}}_{{d^{\rm{Tx}}}'} h_{{d^{\rm{Tx}}}' a} \ell\left({\bf{x}}_{{d^{\rm{Tx}}}'}, {\bf{x}}_{a}\right)$ and $I_{a}^{\rm{C}} = p^{\rm{C}} \sum\limits_{b' \in \left\{\left.{\cal{B}} \right| a\right\}} {\mathbbm{1}}_{b'} h_{b' a} \ell\left({\bf{x}}_{b'}, {\bf{x}}_{a}\right)$ are the sum of the signal powers from D2D transmitter ${d^{\rm{Tx}}}' \in \left\{\left.{\cal{D}}^{\rm{Tx}} \backslash \left\{{d^{\rm{Tx}}}\right\}\right| a\right\}$ and that from BS $b' \in \left\{\left.{\cal{B}} \right| a\right\}$, respectively, and $N_{a}$ is the additive noise at adversary $a$. Therein, $\left\{{\bf{x}}_{{d^{\rm{Tx}}}'}\right\}_{{d^{\rm{Tx}}}' \in \left\{\left.{\cal{D}}^{\rm{Tx}} \backslash \left\{{d^{\rm{Tx}}}\right\}\right| a\right\}} = \Phi_{\left\{\left.{\cal{D}}^{\rm{Tx}} \backslash \left\{{d^{\rm{Tx}}}\right\}\right| a\right\}}$ is the PP of the D2D transmitters that excludes D2D transmitter ${d^{\rm{Tx}}}$ with adversary $a$ as the observation point and $\left\{{\bf{x}}_{b'}\right\}_{b' \in \left\{\left.{\cal{B}} \right| a\right\}} = \Phi_{\left\{\left.{\cal{B}} \right| a\right\}}$ is the PP of the BSs with adversary $a$ as the observation point. Again, similar to~(\ref{eq:pro_dis_b_u}), as no adversary is nearer to D2D transmitter $d^{\rm{Tx}}$ than adversary $a$, the distribution of the distance between D2D transmitter $d^{\rm{Tx}}$ and adversary $a$, i.e., $\left\|{\bf{x}}_{d^{\rm{Tx}}} - {\bf{x}}_{a}\right\|$, is as follows:
\begin{equation}\label{eq:pro_dis_d_TX_a}
f_{\left\|{\bf{x}}_{d^{\rm{Tx}}} - {\bf{x}}_{a}\right\|} \left(r\right) = f_{r_{d^{\rm{Tx}} a}} \left(r\right) = 2 \pi \lambda_{{\cal{A}}} r \exp\left( -\pi \lambda_{{\cal{A}}} r^2\right).
\end{equation} 
Adversary $a$ follows a threshold-based rule to detect the transmission of D2D transmitter $d^{\rm{Tx}}$~\cite{9108996}. In particular, adversary $a$ will advocate ${\cal{D}}_1$ and ${\cal{D}}_0$ when its received signal power is larger than and smaller than a predefined threshold $\tau$, respectively. In this case, adversary $a$ will make the following two types of erroneous decision:
\begin{itemize}
\item False alarm (FA): adversary $a$ will make a FA if its received signal power is larger than $\tau$ while D2D transmitter $d^{\rm{Tx}}$ is mute, and the probability of which is 
\begin{equation}\label{eq:Prob_FA_def}
P^{\rm{FA}}_{a} \left(p^{\rm{D}}, p^{\rm{C}}, \tau\right) = {\mathbb{P}}\left[\left. y_{a} > \tau \right| {\cal{D}}_0 \right].
\end{equation}

\item Miss detection (MD): adversary $a$ will incur a MD if its received signal power is smaller than $\tau$ while D2D transmitter $d^{\rm{Tx}}$ is active, and the probability of which is
\begin{equation}\label{eq:Prob_MD_def}
P^{\rm{MD}}_{a} \left(p^{\rm{D}}, p^{\rm{C}}, \tau\right) = {\mathbb{P}}\left[\left. y_{a} < \tau \right| {\cal{D}}_1 \right].
\end{equation}
\end{itemize}


\subsection{Performance Metric Derivation and Verification}

\subsubsection{Adversary}

For adversary $a$, the FA probability defined in~(\ref{eq:Prob_FA_def}) can be derived as follows:
\begin{equation}
\begin{aligned}
P^{\rm{FA}}_{a} \left(p^{\rm{D}}, p^{\rm{C}}, \tau\right) = & {\mathbb{P}}\left[\left. y_{a} > \tau \right| {\cal{D}}_0 \right] \mathop = \limits^{(\ref{eq:signal_model_a})} {\mathbb{P}}\left[ I_{a}^{\rm{D}} + I_{a}^{\rm{C}} + N_{a} > \tau \right]\\
= & {\mathbb{P}}\left[ I_{a}^{\rm{D}} + I_{a}^{\rm{C}} > \tau - N_{a} \right] = 1 - F_{I_{a}^{\rm{D}} + I_{a}^{\rm{C}}} \left(\tau - N_{a}\right),
\end{aligned}
\end{equation}
where $F_{I_{a}^{\rm{D}} + I_{a}^{\rm{C}}}\left(\cdot\right)$ is the cumulative distribution function (CDF) of the aggregated signal power from the D2D and cellular networks and can be obtained by calculating the inverse Laplace transform of the Laplace transform of $I_{a}^{\rm{D}} + I_{a}^{\rm{C}}$ as follows~\cite{Cohen2007}:
\begin{enumerate}
\item Laplace transform of $I_{a}^{\rm{D}} + I_{a}^{\rm{C}}$, i.e., ${\cal{L}}_{I_{a}^{\rm{D}} + I_{a}^{\rm{C}}}\left(s\right)$: 
\begin{equation}
\begin{aligned}
{\cal{L}}_{I_{a}^{\rm{D}} + I_{a}^{\rm{C}}}\left(s\right) = & {\mathbb{E}}_{I_{a}^{\rm{D}} + I_{a}^{\rm{C}}}\left[\exp\left(-s \left(I_{a}^{\rm{D}} + I_{a}^{\rm{C}}\right)\right)\right] \\
\mathop = \limits ^{(a)} & {\mathbb{E}}_{I_{a}^{\rm{D}}}\left[\exp\left(-s I_{a}^{\rm{D}}\right)\right] {\mathbb{E}}_{I_{a}^{\rm{C}}}\left[\exp\left(-s I_{a}^{\rm{C}}\right)\right],
\end{aligned}
\end{equation}
where $(a)$ follows the independence between $I_{a}^{\rm{D}}$ and $I_{a}^{\rm{C}}$,
\begin{equation}\label{eq:laplace_transform_D2D_interference_a}
\begin{aligned}
& {\mathbb{E}}_{I_{a}^{\rm{D}}}\left[\exp\left(-s I_{a}^{\rm{D}}\right)\right] \\
= & {\mathbb{E}}_{\Phi_{\left\{\left.{\cal{D}}^{\rm{Tx}} \backslash \left\{{d^{\rm{Tx}}}\right\}\right| a\right\}}}\left[\prod\limits_{{d^{\rm{Tx}}}' \in \left\{\left.{\cal{D}}^{\rm{Tx}} \backslash \left\{{d^{\rm{Tx}}}\right\}\right| a\right\}} {\mathbb{E}}_{h_{{d^{\rm{Tx}}}' a}} \left[\exp\left(-s p^{\rm{D}} {\mathbbm{1}}_{{d^{\rm{Tx}}}'} h_{{d^{\rm{Tx}}}' a} \ell\left({\bf{x}}_{{d^{\rm{Tx}}}'}, {\bf{x}}_{a}\right)\right)\right]\right]\\
= & {\mathbb{E}}_{\Phi_{\left\{\left.{\cal{D}}^{\rm{Tx}} \backslash \left\{{d^{\rm{Tx}}}\right\}\right| a\right\}}}\left[\prod\limits_{{d^{\rm{Tx}}}' \in \left\{\left.{\cal{D}}^{\rm{Tx}} \backslash \left\{{d^{\rm{Tx}}}\right\}\right| a\right\}} \left(\frac{{\mathbb{P}}^{{\cal{D}}_1}}{s p^{\rm{D}} \ell\left({\bf{x}}_{{d^{\rm{Tx}}}'}, {\bf{x}}_{a}\right) + 1} + 1 - {\mathbb{P}}^{{\cal{D}}_1}\right)\right]\\
= & {\mathbb{E}}_{\Phi_{\left\{\left.{\cal{D}}^{\rm{Tx}} \backslash \left\{{d^{\rm{Tx}}}\right\}\right| a\right\}}}\left[\prod\limits_{{d^{\rm{Tx}}}' \in \left\{\left.{\cal{D}}^{\rm{Tx}} \backslash \left\{{d^{\rm{Tx}}}\right\}\right| a\right\}} \left(1 - \frac{s {\mathbb{P}}^{{\cal{D}}_1} p^{\rm{D}} \ell\left({\bf{x}}_{{d^{\rm{Tx}}}'}, {\bf{x}}_{a}\right)}{s p^{\rm{D}} \ell\left({\bf{x}}_{{d^{\rm{Tx}}}'}, {\bf{x}}_{a}\right) + 1}\right)\right]\\
\mathop = \limits^{(b)}& \exp\left( - 2 \pi \lambda_{{\cal{D}}^{\rm{Tx}}} \int^\infty_0 \frac{s {\mathbb{P}}^{{\cal{D}}_1} p^{\rm{D}} r_{{d^{\rm{Tx}}}' a}^{-\alpha}}{s p^{\rm{D}} r_{{d^{\rm{Tx}}}' a}^{-\alpha} + 1} r_{{d^{\rm{Tx}}}' a} {\rm{d}} r_{{d^{\rm{Tx}}}' a} \right)\\
= & \exp\left( - 2 \pi \lambda_{{\cal{D}}^{\rm{Tx}}} {\mathbb{P}}^{{\cal{D}}_1} \int^\infty_0 \frac{s p^{\rm{D}} r_{{d^{\rm{Tx}}}' a}}{s p^{\rm{D}} + r_{{d^{\rm{Tx}}}' a}^{\alpha}} {\rm{d}} r_{{d^{\rm{Tx}}}' a} \right)\\
= & \exp\left( - 2 \pi \lambda_{{\cal{D}}^{\rm{Tx}}} {\mathbb{P}}^{{\cal{D}}_1} \left(s p^{\rm{D}}\right)^{\frac{2}{\alpha}}\int^\infty_0 \frac{r}{1 + r^{\alpha}} {\rm{d}} r \right) = \exp\left( - \frac{\pi \lambda_{{\cal{D}}^{\rm{Tx}}} {\mathbb{P}}^{{\cal{D}}_1} \left(s p^{\rm{D}}\right)^{\frac{2}{\alpha}}}{{\rm{sinc}}\left(\frac{2}{\alpha}\right)}\right)
\end{aligned}
\end{equation}
with (b) following the probability generating functionals of PPP~\cite{9736993}, and
\begin{equation}
{\mathbb{E}}_{I_{a}^{\rm{C}}}\left[\exp\left(-s I_{a}^{\rm{C}}\right)\right] \mathop = \limits^{(\ref{eq:laplace_transform_D2D_interference_a})} \exp\left( - \frac{\pi \lambda_{{\cal{B}}} {\mathbb{P}}^{{\cal{C}}_1} \left(s p^{\rm{C}}\right)^{\frac{2}{\alpha}}}{{\rm{sinc}}\left(\frac{2}{\alpha}\right)}\right).
\end{equation}

\item Inverse Laplace transform of ${\cal{L}}_{I_{a}^{\rm{D}} + I_{a}^{\rm{C}}}\left(s\right)$, i.e., ${\cal{L}}^{-1}_{{\cal{L}}_{I_{a}^{\rm{D}} + I_{a}^{\rm{C}}}\left(s\right)} \left(t\right)$: according to Chapter~2 of~\cite{Cohen2007},
\begin{equation}\label{eq:prob_I_D_a_I_C_a}
\begin{aligned}
& F_{I_{a}^{\rm{D}} + I_{a}^{\rm{C}}} \left(t\right) \triangleq {\cal{L}}^{-1}_{{\cal{L}}_{I_{a}^{\rm{D}} + I_{a}^{\rm{C}}}\left(s\right)} \left(t\right) \\
= & 1 - \int^\infty_0 \frac{1}{\pi v} \exp\left(-v^{\frac{2}{\alpha}} \iota_{I_{a}^{\rm{D}} + I_{a}^{\rm{C}}} \cos\left(\frac{2\pi}{\alpha}\right) - tv\right) \sin \left(\sin\left(\frac{2\pi}{\alpha}\right) \iota_{I_{a}^{\rm{D}} + I_{a}^{\rm{C}}} v^{\frac{2}{\alpha}}\right) {\rm{d}} v,
\end{aligned}
\end{equation}
where $\iota_{I_{a}^{\rm{D}} + I_{a}^{\rm{C}}} = \frac{\pi}{{\rm{sinc}}\left(\frac{2}{\alpha}\right)}\left(\lambda_{{\cal{B}}} {\mathbb{P}}^{{\cal{C}}_1} \left(p^{\rm{C}}\right)^{\frac{2}{\alpha}} + \lambda_{{\cal{D}}^{\rm{Tx}}} {\mathbb{P}}^{{\cal{D}}_1} \left(p^{\rm{D}}\right)^{\frac{2}{\alpha}}\right)$.

\end{enumerate}
The MD probability of adversary $a$ defined in~(\ref{eq:Prob_MD_def}) can be derived as follows:
\begin{equation}
\begin{aligned}
P^{\rm{MD}}_{a} \left(p^{\rm{D}}, p^{\rm{C}}, \tau\right) = & {\mathbb{P}}\left[\left. y_{a} < \tau \right| {\cal{D}}_1 \right]
\mathop = \limits^{(\ref{eq:signal_model_a})} {\mathbb{P}}\left[ p^{\rm{D}} h_{{d^{\rm{Tx}}} a} \ell\left({\bf{x}}_{d^{\rm{Tx}}}, {\bf{x}}_{a}\right) + I_{a}^{\rm{D}} + I_{a}^{\rm{C}} + N_{a} < \tau \right]\\
= & \int_0^{\tau - N_{a}} f_{p^{\rm{D}} h_{{d^{\rm{Tx}}} a} \ell\left({\bf{x}}_{d^{\rm{Tx}}}, {\bf{x}}_{a}\right)} \left(t\right) F_{I_{a}^{\rm{D}} + I_{a}^{\rm{C}}} \left(\tau - N_{a} - t\right) {\rm{d}} t,
\end{aligned}
\end{equation}
where $F_{I_{a}^{\rm{D}} + I_{a}^{\rm{C}}} \left(\cdot\right)$ has been shown in~(\ref{eq:prob_I_D_a_I_C_a}) and 
\begin{equation}
\begin{aligned}
& f_{p^{\rm{D}} h_{{d^{\rm{Tx}}} a} \ell\left({\bf{x}}_{d^{\rm{Tx}}}, {\bf{x}}_{a}\right)} \left(t\right) = \frac{{\rm{d}}}{{\rm{d}} t} F_{p^{\rm{D}} h_{{d^{\rm{Tx}}} a} \ell\left({\bf{x}}_{d^{\rm{Tx}}}, {\bf{x}}_{a}\right)} \left(t\right) = \frac{{\rm{d}}}{{\rm{d}} t} {\mathbb{P}}\left[p^{\rm{D}} h_{{d^{\rm{Tx}}} a} \ell\left({\bf{x}}_{d^{\rm{Tx}}}, {\bf{x}}_{a}\right) < t \right]\\
= & \frac{{\rm{d}}}{{\rm{d}} t} {\mathbb{P}}\left[ h_{{d^{\rm{Tx}}} a} < \frac{t r_{{d^{\rm{Tx}}} a}^\alpha}{p^{\rm{D}}} \right] = \frac{{\rm{d}}}{{\rm{d}} t} {\mathbb{E}}_{r_{{d^{\rm{Tx}}} a}}\left[ 1 - \exp\left( -\frac{t r_{{d^{\rm{Tx}}} a}^\alpha}{p^{\rm{D}}} \right)\right]\\
= & \frac{{\rm{d}}}{{\rm{d}} t} \int_0^\infty \left[ 1 - \exp\left( -\frac{t r^\alpha}{p^{\rm{D}}} \right)\right] f_{r_{{d^{\rm{Tx}}} a}} \left(r\right) {\rm{d}} r \mathop = \limits^{(a)} \int_0^\infty \exp\left( -\frac{t r^\alpha}{p^{\rm{D}}} \right) \frac{r^\alpha}{p^{\rm{D}}} f_{r_{{d^{\rm{Tx}}} a}} \left(r\right) {\rm{d}} r.
\end{aligned}
\end{equation}
Therein, $(a)$ follows the Leibniz integral rule to take the differentiation under the integral sign, and $f_{r_{{d^{\rm{Tx}}} a}} \left(\cdot\right)$ is given in~(\ref{eq:pro_dis_d_TX_a}).
The derivation of the FA and MD probabilities can be verified by the comparison between the simulation and analytic results regarding the detection error defined in~(\ref{eq:lower_problem}) shown in Figs.~\ref{fig:simulation_vs_analysis_detection_error_uniqueness_sol_adversary}(a) and (b). 

\subsubsection{D2D Network}

\begin{figure}
	\centering
	\includegraphics[width=0.5\textwidth,trim=10 40 30 70,clip]{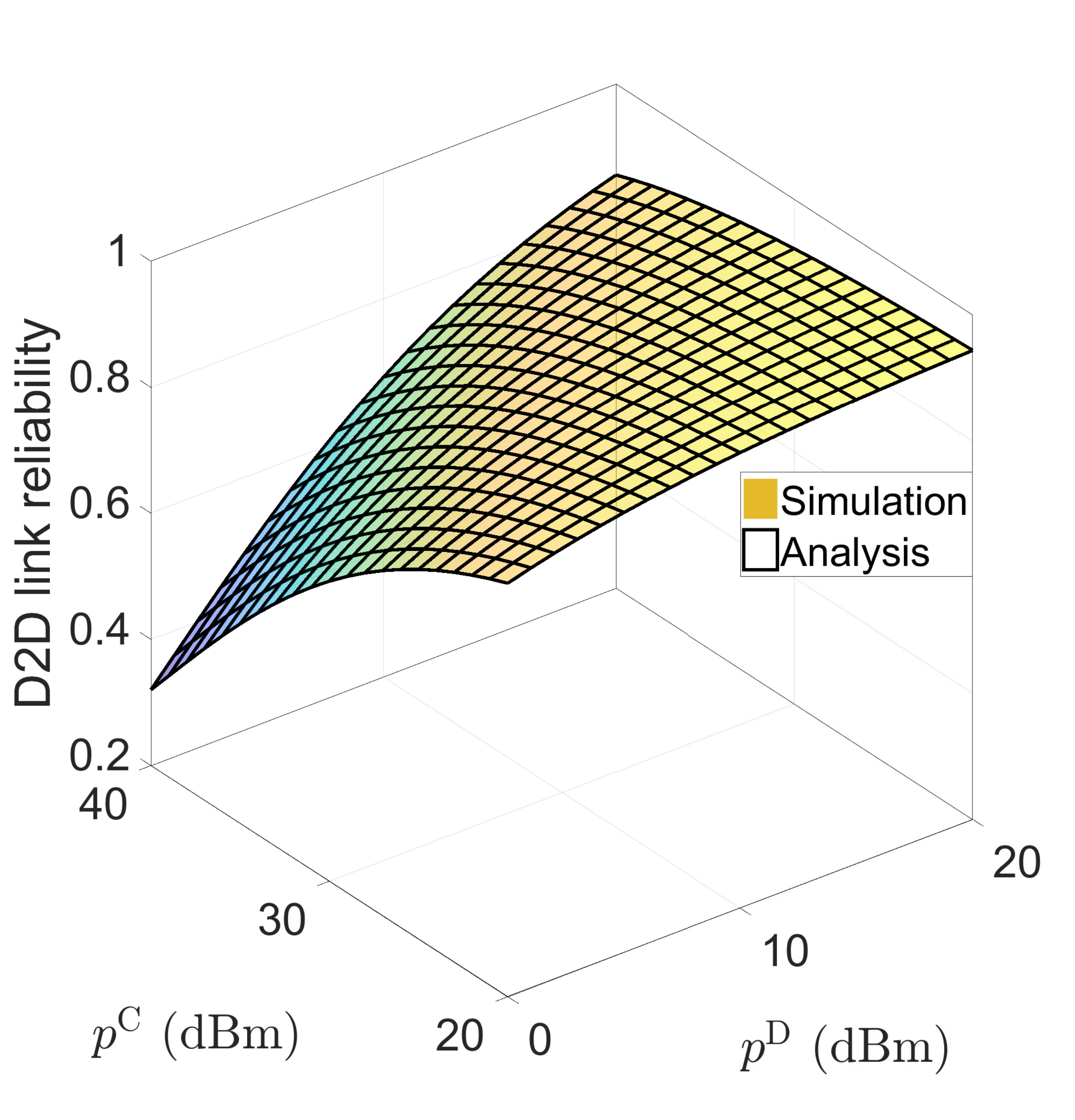}
	\caption{Comparison between the simulation and analytic results on D2D link reliability.}
    \label{fig:simulation_vs_analysis_D2D_reliability}
\end{figure}

For D2D receiver $d^{\rm{Rx}}$, its successful decoding probability defined in~(\ref{eq:Prob_success_D2D}) is  
\begin{equation}
\begin{aligned}
& {\mathbb{P}}\left[\left. {\text{SINR}}_{d^{\rm{Rx}}} > \theta^{\rm{D}} \right| {\cal{D}}_1 \right] = {\mathbb{P}}\left[ h_{d^{\rm{Tx}} d^{\rm{Rx}}} > \frac{\theta^{\rm{D}} R^{\alpha} \left(I_{d^{\rm{Rx}}}^{\rm{D}} + I_{d^{\rm{Rx}}}^{\rm{C}} + N_{d^{\rm{Rx}}}\right)}{p^{\rm{D}}}\right]\\
= & \exp\left(-\frac{\theta^{\rm{D}} R^{\alpha} N_{d^{\rm{Rx}}}}{p^{\rm{D}}}\right) {\mathbb{E}}_{I_{d^{\rm{Rx}}}^{\rm{D}} + I_{d^{\rm{Rx}}}^{\rm{C}}}\left[ \exp\left(-\frac{\theta^{\rm{D}} R^{\alpha} \left(I_{d^{\rm{Rx}}}^{\rm{D}} + I_{d^{\rm{Rx}}}^{\rm{C}} \right)}{p^{\rm{D}}}\right)\right]\\
\mathop = \limits^{(\ref{eq:laplace_transform_D2D_interference_a})} & \exp\left(-\frac{\theta^{\rm{D}} R^{\alpha} N_{d^{\rm{Rx}}}}{p^{\rm{D}}}\right) \exp\left(-\frac{\pi \left(\theta^{\rm{D}}\right)^{\frac{2}{\alpha}} R^2 }{{\rm{sinc}}\left(\frac{2}{\alpha}\right)} \left(\lambda_{{\cal{D}}^{\rm{Tx}}} {\mathbb{P}}^{{\cal{D}}_1} + \lambda_{{\cal{B}}} {\mathbb{P}}^{{\cal{C}}_1} \left(\frac{p^{\rm{C}}}{p^{\rm{D}}}\right)^{\frac{2}{\alpha}}\right) \right),
\end{aligned}
\end{equation}
which can be verified by the consistency between the simulation and analytic results as shown in Fig.~\ref{fig:simulation_vs_analysis_D2D_reliability}.

\subsubsection{Cellular Network}

\begin{figure}
     \centering
     \begin{minipage}{8cm}
		\centering
		\includegraphics[width=1\textwidth,trim=10 40 30 30,clip]{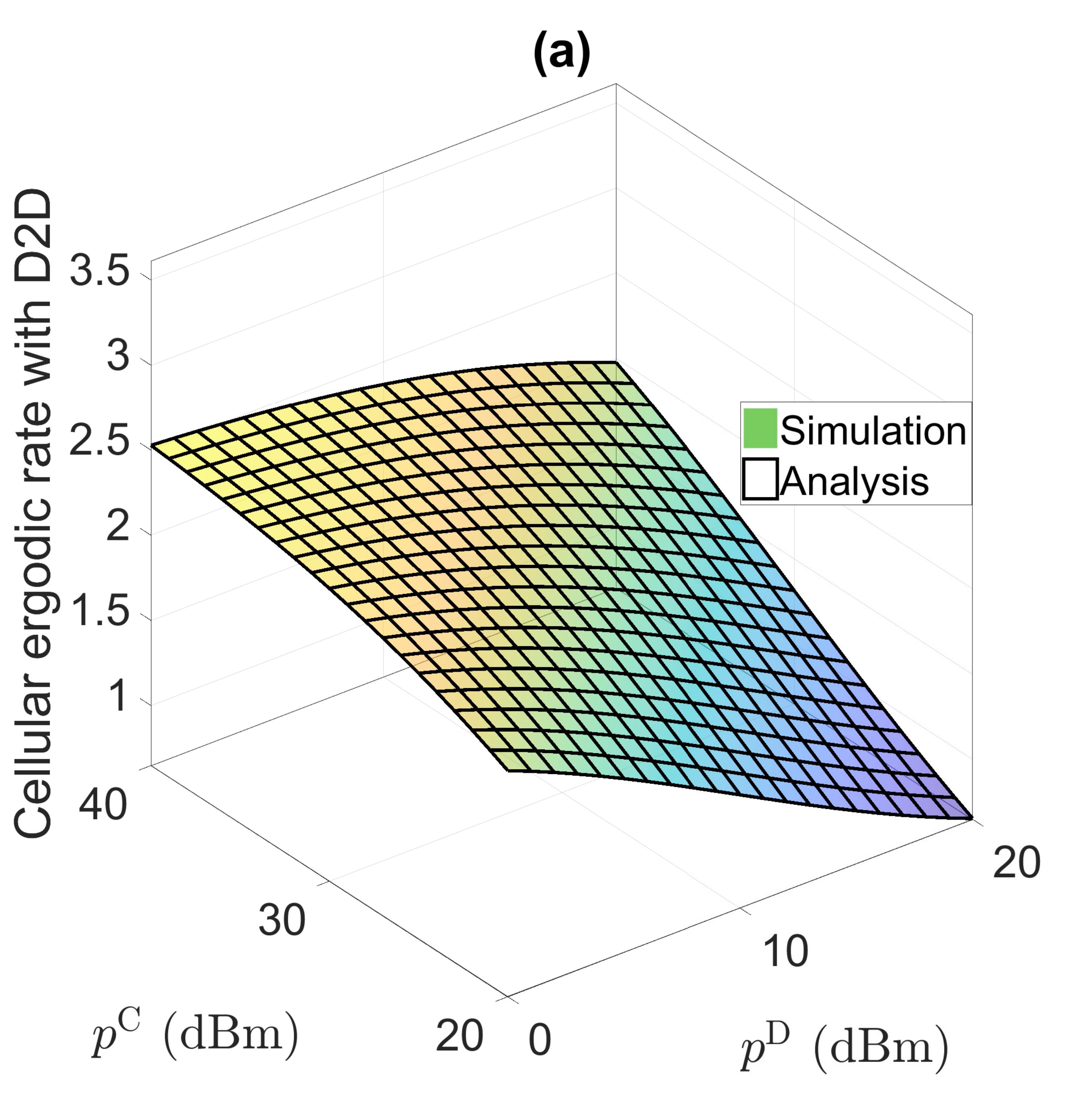}
     \end{minipage}
     \begin{minipage}{8cm}
		\centering
		\includegraphics[width=1\textwidth,trim=10 40 30 30,clip]{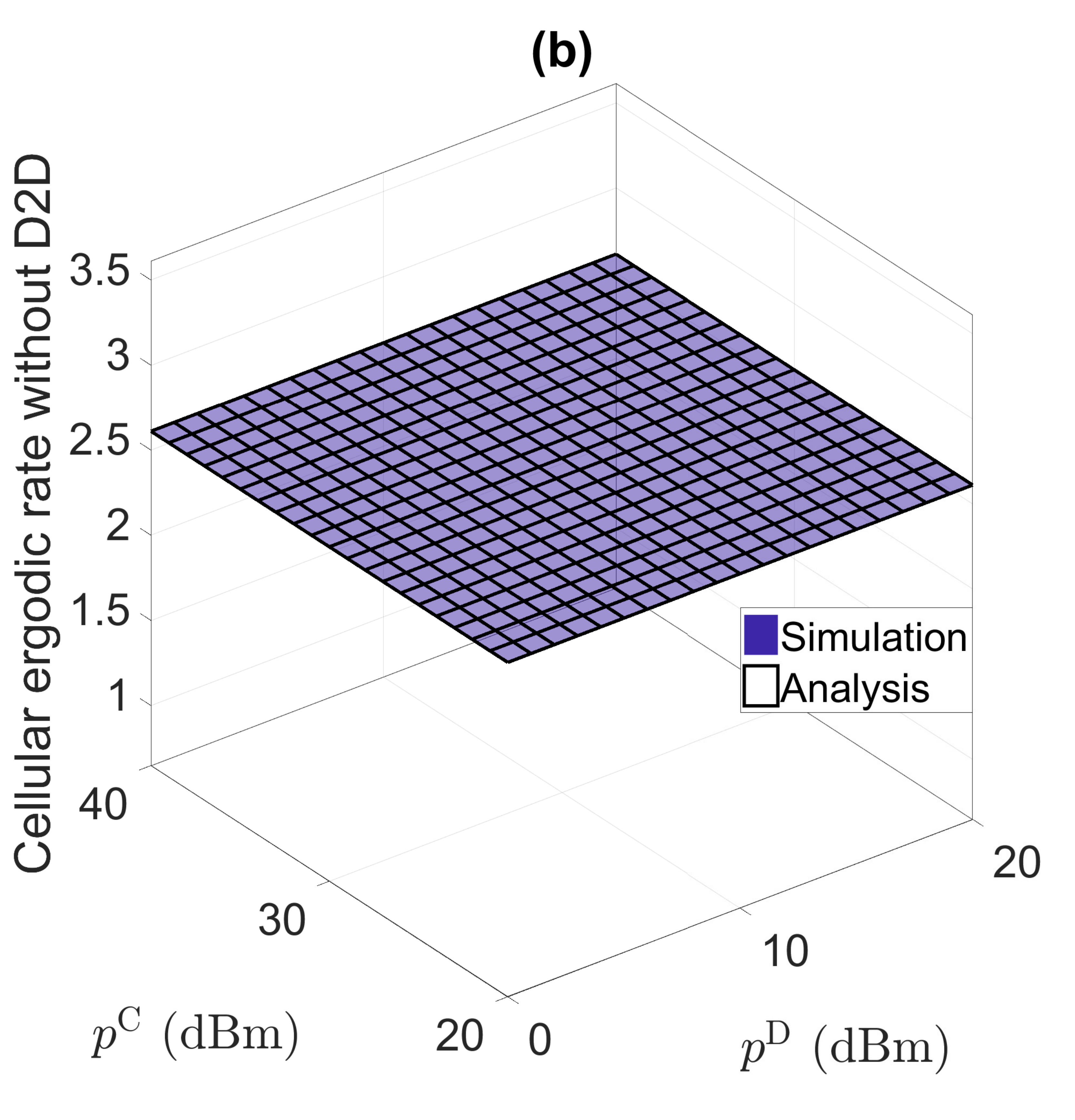}
     \end{minipage}
        \caption{Comparison between the simulation and analytic results on cellular ergodic rate.}
        \label{fig:simulation_vs_analysis_ergodic_rate}
\end{figure}

Based on the signal model defined in~(\ref{eq:received_signal_u_with_D2D}), the successful decoding probability at CU $u$ taking into account the interference from the D2D network is 
\begin{equation}\label{eq:Prob_success_cellular_w_D2D_specific}
\begin{aligned}
& {\mathbb{P}}\left[\left. {\text{SINR}}_{u} > \theta^{\rm{C}} \right| {\cal{C}}_1\right] = {\mathbb{P}}\left[ \frac{p^{\rm{C}} h_{b u} \ell\left({\bf{x}}_{b}, {\bf{x}}_{u}\right)} {I_{u}^{\rm{D}} + I_{u}^{\rm{C}} + N_{u}} > \theta^{\rm{C}} \right]\\
= & {\mathbb{P}}\left[ h_{b u} > \frac{\theta^{\rm{C}}\left(I_{u}^{\rm{D}} + I_{u}^{\rm{C}} + N_{u}\right)} {p^{\rm{C}} \ell\left({\bf{x}}_{b}, {\bf{x}}_{u}\right)} \right] = {\mathbb{E}}\left[ \exp\left( -\frac{\theta^{\rm{C}}\left(I_{u}^{\rm{D}} + I_{u}^{\rm{C}} + N_{u}\right)} {p^{\rm{C}} \ell\left({\bf{x}}_{b}, {\bf{x}}_{u}\right)} \right) \right]\\
= & {\mathbb{E}}_{r_{b u}}\left[ \exp\left( -\frac{\theta^{\rm{C}} r_{b u}^\alpha N_{u}} {p^{\rm{C}}} \right) {\mathbb{E}}_{I_{u}^{\rm{D}}}\left[\exp\left( -\frac{\theta^{\rm{C}} r_{b u}^\alpha I_{u}^{\rm{D}}} {p^{\rm{C}}} \right)\right] {\mathbb{E}}_{I_{u}^{\rm{C}}}\left[\exp\left( -\frac{\theta^{\rm{C}} r_{b u}^\alpha I_{u}^{\rm{C}}} {p^{\rm{C}}} \right) \right]\right],
\end{aligned}
\end{equation}
where ${\mathbb{E}}_{I_{u}^{\rm{D}}}\left[\exp\left( -\frac{\theta^{\rm{C}} r_{b u}^\alpha I_{u}^{\rm{D}}} {p^{\rm{C}}} \right)\right] \mathop = \limits^{(\ref{eq:laplace_transform_D2D_interference_a})} \exp\left( - \frac{\pi \lambda_{{\cal{D}}^{\rm{Tx}}} {\mathbb{P}}^{{\cal{D}}_1} r_{b u}^2 \left(\theta^{\rm{C}}\frac{ p^{\rm{D}}}{p^{\rm{C}}}\right)^{\frac{2}{\alpha}}}{{\rm{sinc}}\left(\frac{2}{\alpha}\right)}\right)$, the probability density function (PDF) of $r_{b u}$, i.e., $f_{r_{b u}}\left(\cdot\right)$, is given in~(\ref{eq:pro_dis_b_u}), and 
\begin{equation}\label{eq:Prob_success_cellular_w_D2D_specific_third_term}
\begin{aligned}
&{\mathbb{E}}_{I_{u}^{\rm{C}}}\left[\exp\left( -\frac{\theta^{\rm{C}} r_{b u}^\alpha I_{u}^{\rm{C}}} {p^{\rm{C}}} \right) \right]\\
=& {\mathbb{E}}_{\Phi_{\left\{\left.{\cal{B}} \backslash \left\{b\right\}\right| u\right\}}}\left[\prod\limits_{b' \in \left\{\left.{\cal{B}} \backslash \left\{b\right\}\right| u\right\}} {\mathbb{E}}_{h_{b' u}}\left[\exp\left( -\frac{\theta^{\rm{C}} r_{b u}^\alpha} {p^{\rm{C}}} {\mathbbm{1}}_{b'} p^{\rm{C}} h_{b' u} \ell\left({\bf{x}}_{b'}, {\bf{x}}_{u}\right) \right) \right]\right]\\
=& {\mathbb{E}}_{\Phi_{\left\{\left.{\cal{B}} \backslash \left\{b\right\}\right| u\right\}}}\left[\prod\limits_{b' \in \left\{\left.{\cal{B}} \backslash \left\{b\right\}\right| u\right\}} \left( \frac{{\mathbb{P}}^{{\cal{C}}_1}}{\theta^{\rm{C}} r_{b u}^\alpha r_{b' u}^{-\alpha} + 1} + 1 - {\mathbb{P}}^{{\cal{C}}_1}\right)\right]\\
=& \exp\left(-2\pi\lambda_{{\cal{B}}}\int^\infty_{r_{bu}} \frac{{\mathbb{P}}^{{\cal{C}}_1} \theta^{\rm{C}} r_{b u}^\alpha}{\theta^{\rm{C}} r_{b u}^\alpha + r_{b' u}^{\alpha}} r_{b' u} {\rm{d}} r_{b' u} \right)\\=& \exp\left(-2 \pi \lambda_{{\cal{B}}} {\mathbb{P}}^{{\cal{C}}_1} r_{b u}^2 \left(\theta^{\rm{C}}\right)^{\frac{2}{\alpha}} \int^\infty_{\left(\theta^{\rm{C}}\right)^{-\frac{1}{\alpha}}} \frac{u}{1 + u^{\alpha}} {\rm{d}} u \right).
\end{aligned}
\end{equation}
Accordingly, the ergodic rate of CU $u$ taking into account the interference from the D2D network is~\cite{7497569}
\begin{equation}\label{eq:ergodic_rate_u_with_D2D}
\begin{aligned}
& {\overline{R}}_u\left(p^{\rm{D}}, p^{\rm{C}}\right) = \int\limits_0^\infty {\mathbb{P}}\left[\left. \log_2\left(1 + {\text{SINR}}_{u} \right) > R^{\rm{C}} \right| {\cal{C}}_1\right] {\rm{d}} R^{\rm{C}} \\
= & \int\limits_0^\infty {\mathbb{P}}\left[\left. {\text{SINR}}_{u} > 2^{R^{\rm{C}}} - 1 \right| {\cal{C}}_1\right] {\rm{d}} R^{\rm{C}} = \int\limits_0^\infty \left.{\mathbb{P}}\left[\left. {\text{SINR}}_{u} >  \theta^{\rm{C}} \right| {\cal{C}}_1\right]\right|_{\theta^{\rm{C}} = 2^{R^{\rm{C}}} - 1} {\rm{d}} R^{\rm{C}},
\end{aligned}
\end{equation}
where ${\mathbb{P}}\left[\left. {\text{SINR}}_{u} >  \theta^{\rm{C}} \right| {\cal{C}}_1\right]$ is given in~(\ref{eq:Prob_success_cellular_w_D2D_specific}). The derivation of ${\overline{R}}_u\left(p^{\rm{D}}, p^{\rm{C}}\right)$ can be verified by the consistency between the simulation and analytic results shown in Fig.~\ref{fig:simulation_vs_analysis_ergodic_rate}(a).

Similar to~(\ref{eq:Prob_success_cellular_w_D2D_specific})-(\ref{eq:ergodic_rate_u_with_D2D}) and based on~(\ref{eq:received_signal_u_without_D2D}), the successful decoding probability and the ergodic rate of CU $u$ without taking into account the interference from the D2D network are 
\begin{equation}\label{eq:Prob_success_cellular_wo_D2D_specific}
{\mathbb{P}}\left[\left. {\widetilde{\text{SINR}}}_{u} > \theta^{\rm{C}} \right| {\cal{C}}_1\right] = {\mathbb{E}}_{r_{b u}}\left[ \exp\left( -\frac{\theta^{\rm{C}} r_{b u}^\alpha N_{u}} {p^{\rm{C}}} \right) {\mathbb{E}}_{I_{u}^{\rm{C}}}\left[\exp\left( -\frac{\theta^{\rm{C}} r_{b u}^\alpha I_{u}^{\rm{C}}} {p^{\rm{C}}} \right) \right] \right]
\end{equation}
and
\begin{equation}\label{eq:ergodic_rate_u_wo_D2D}
\begin{aligned}
& {\overline{\widetilde{R}}}_u\left(p^{\rm{D}}, p^{\rm{C}}\right) = \int\limits_0^\infty {\mathbb{P}}\left[\left. \log_2\left(1 + {\widetilde{\text{SINR}}}_{u} \right) > R^{\rm{C}} \right| {\cal{C}}_1\right] {\rm{d}} R^{\rm{C}} \\
= & \int\limits_0^\infty {\mathbb{P}}\left[\left. {\widetilde{\text{SINR}}}_{u} > 2^{R^{\rm{C}}} - 1 \right| {\cal{C}}_1\right] {\rm{d}} R^{\rm{C}} \mathop = \limits^{(\ref{eq:Prob_success_cellular_wo_D2D_specific})} \int\limits_0^\infty \left.{\mathbb{P}}\left[\left. {\widetilde{\text{SINR}}}_{u} > \theta^{\rm{C}}\right| {\cal{C}}_1\right] \right|_{\theta^{\rm{C}} = 2^{R^{\rm{C}}} - 1}{\rm{d}} R^{\rm{C}}, 
\end{aligned}
\end{equation}
respectively, where ${\mathbb{E}}_{I_{u}^{\rm{C}}}\left[\exp\left( -\frac{\theta^{\rm{C}} r_{b u}^\alpha I_{u}^{\rm{C}}} {p^{\rm{C}}} \right) \right]$ and the PDF of $r_{b u}$, i.e., $f_{r_{b u}}\left(\cdot\right)$, of~(\ref{eq:Prob_success_cellular_wo_D2D_specific}) are given in~(\ref{eq:Prob_success_cellular_w_D2D_specific_third_term}) and~(\ref{eq:pro_dis_b_u}), respectively. The derivation of ${\overline{\widetilde{R}}}_u\left(p^{\rm{D}}, p^{\rm{C}}\right)$ can be verified by Fig.~\ref{fig:simulation_vs_analysis_ergodic_rate}(b). 

The expectation of the interference from the D2D network to CU $u$ is 
\begin{equation}\label{eq:expectation_I_D_u}
{\mathbb{E}}\left[I_{u}^{\rm{D}}\right] = \int^\infty_0 1 - F_{I_{u}^{\rm{D}}} \left(t\right) {\rm{d}} t,
\end{equation}
where
\begin{equation}
F_{I_{u}^{\rm{D}}} \left(t\right) \mathop = \limits^{(\ref{eq:prob_I_D_a_I_C_a})} 1 - \int^\infty_0 \frac{1}{\pi v} \exp\left(-v^{\frac{2}{\alpha}} \iota_{I_{u}^{\rm{D}}} \cos\left(\frac{2\pi}{\alpha}\right) - tv\right) \sin \left(\sin\left(\frac{2\pi}{\alpha}\right) \iota_{I_{u}^{\rm{D}}} v^{\frac{2}{\alpha}}\right) {\rm{d}} v
\end{equation}
with $\iota_{I_{u}^{\rm{D}}} =  - \frac{\lambda_{{\cal{D}}^{\rm{Tx}}} \pi {\mathbb{P}}^{{\cal{D}}_1} \left(p^{\rm{D}}\right)^{\frac{2}{\alpha}}}{{\rm{sinc}}\left(\frac{2}{\alpha}\right)}$. Here, due to the consistency between the simulation and analytic results over the cellular ergodic rate as shown in Fig.~\ref{fig:simulation_vs_analysis_ergodic_rate}, the numerical verification of the derivation of ${\mathbb{E}}\left[I_{u}^{\rm{D}}\right]$ becomes unnecessary and hence is omitted here.


\section{Problem Statement and Solution Analysis}
\label{sec:formulation_analysis}

We model the combat between the adversary and the legitimate entity, i.e., the D2D-underlaid cellular network, in order to capture the conflict of interests between them. As stated in Section~\ref{sec:system_model}, the adversary aims to detect the D2D communication based on its observation, i.e., its received signal power from the legitimate entity. In this circumstance, the decision-making of the adversary is after that of the legitimate entity, which induces a bi-level decision-making framework for the combat. This can be captured by a two-stage Stackelberg game framework with the legitimate entity as the leader at the upper stage deciding the cellular and D2D transmission powers and the adversary as the follower at the lower stage deciding the power detection threshold. The two-stage Stackelberg game formulation regarding the representative network nodes, i.e., D2D transmitter $d^{\rm{Tx}}$, D2D receiver $d^{\rm{Rx}}$, BS $b$, CU $u$, and adversary $a$, is a single-follower-single-leader game and presented together with the solution analysis for each stage as follows.


\subsection{Lower-stage Detection Error Minimization Problem Formulation and Solution Analysis}

\subsubsection{Problem Formulation}

At the lower stage, given the D2D and cellular transmission powers, i.e., $p^{\rm{D}}$ and $p^{\rm{C}}$, respectively, adversary $a$ aims to minimize its detection error probability based on its received signal power, i.e., $\min\limits_{\tau} \left(1 - {\mathbb{P}}^{{\cal{D}}_1}\right)P^{\rm{FA}}_{a} \left(p^{\rm{D}}, p^{\rm{C}}, \tau\right) + {\mathbb{P}}^{{\cal{D}}_1} P^{\rm{MD}}_{a} \left(p^{\rm{D}}, p^{\rm{C}}, \tau\right)$, where $P^{\rm{FA}}_{a} \left(p^{\rm{D}}, p^{\rm{C}}, \tau\right)$ and $P^{\rm{MD}}_{a} \left(p^{\rm{D}}, p^{\rm{C}}, \tau\right)$ are defined in~(\ref{eq:Prob_FA_def}) and~(\ref{eq:Prob_MD_def}), respectively. However, as adversary $a$ is unaware of the activation status of D2D transmitter $d^{\rm{Tx}}$, i.e., ${\mathbb{P}}^{{\cal{D}}_1}$, we consider that adversary $a$ aims to minimize the upper bound of its detection error probability as follows:
\begin{equation}\label{eq:lower_problem}
\begin{aligned}
&\left(1 - {\mathbb{P}}^{{\cal{D}}_1}\right)P^{\rm{FA}}_{a} \left(p^{\rm{D}}, p^{\rm{C}}, \tau\right) + {\mathbb{P}}^{{\cal{D}}_1} P^{\rm{MD}}_{a} \left(p^{\rm{D}}, p^{\rm{C}}, \tau\right) \\
\le & \max\left\{\left(1 - {\mathbb{P}}^{{\cal{D}}_1}\right), {\mathbb{P}}^{{\cal{D}}_1}\right\} \left[P^{\rm{FA}}_{a} \left(p^{\rm{D}}, p^{\rm{C}}, \tau\right) + P^{\rm{MD}}_{a} \left(p^{\rm{D}}, p^{\rm{C}}, \tau\right)\right]\\
\Rightarrow \tau^\star = & \arg\min_{\tau} P^{\rm{FA}}_{a} \left(p^{\rm{D}}, p^{\rm{C}}, \tau\right) + P^{\rm{MD}}_{a} \left(p^{\rm{D}}, p^{\rm{C}}, \tau\right),
\end{aligned}
\end{equation}
where $P^{\rm{FA}}_{a} \left(p^{\rm{D}}, p^{\rm{C}}, \tau\right) + P^{\rm{MD}}_{a} \left(p^{\rm{D}}, p^{\rm{C}}, \tau\right)$ is defined as the detection error in this paper.

\subsubsection{Solution Analysis} 

\begin{figure}
     \centering
     \begin{minipage}{5cm}
		\centering
		\includegraphics[width=1\textwidth,trim=10 0 15 10,clip]{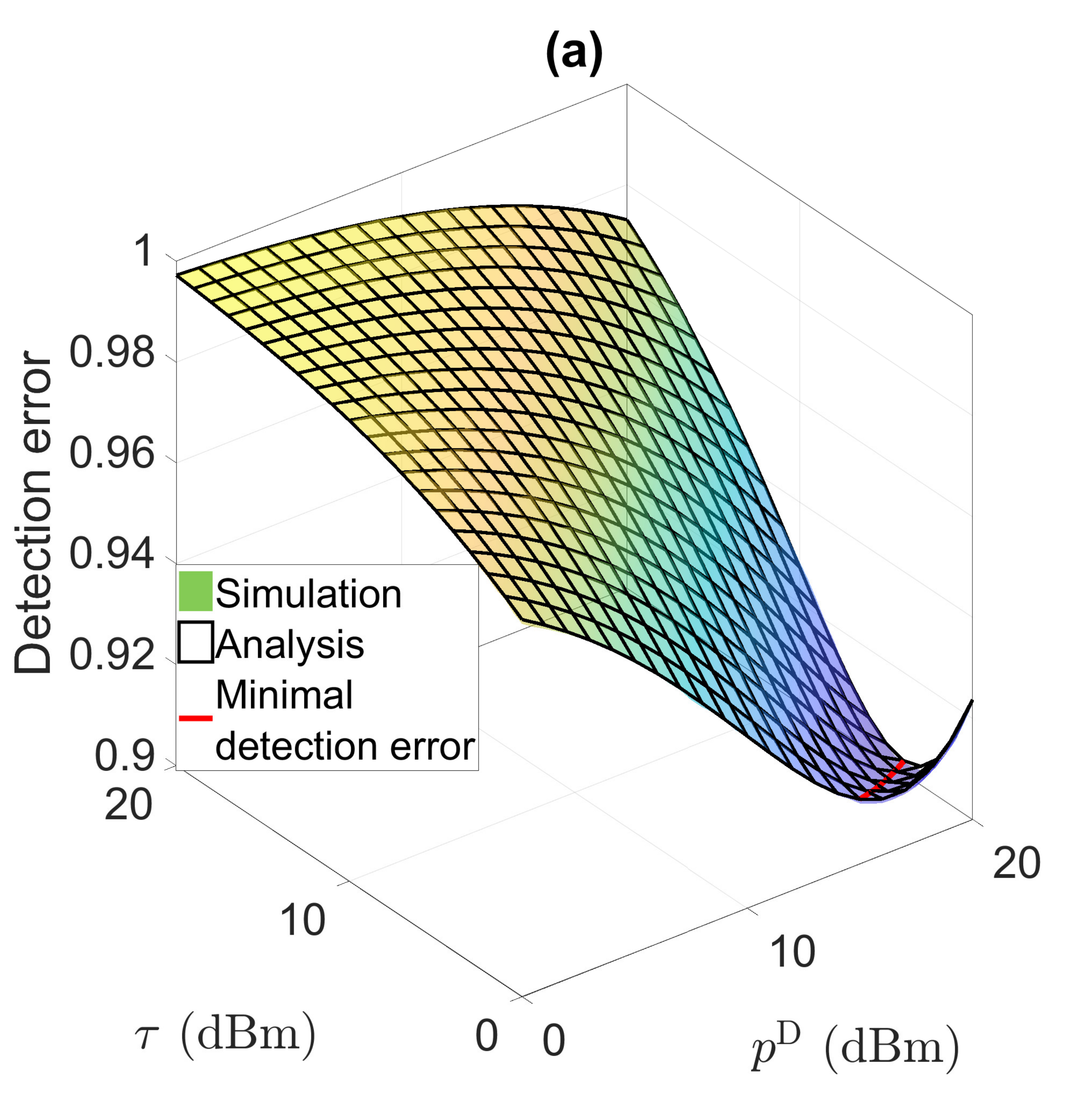}
     \end{minipage}
     \begin{minipage}{5cm}
		\centering
		\includegraphics[width=1\textwidth,trim=10 0 15 10,clip]{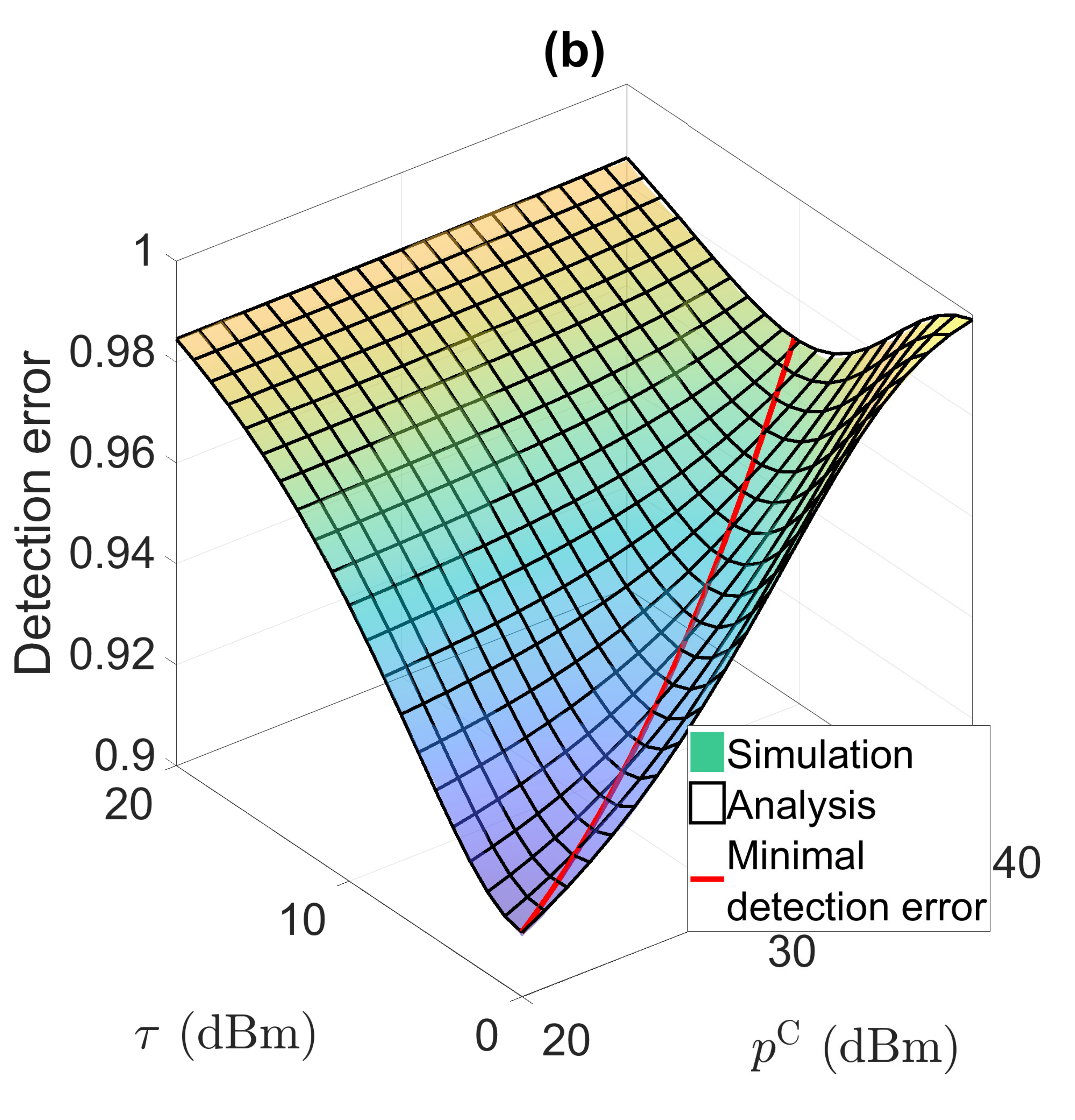}
     \end{minipage}
     \begin{minipage}{5cm}
		\centering
		\includegraphics[width=1\textwidth,trim=10 0 15 10,clip]{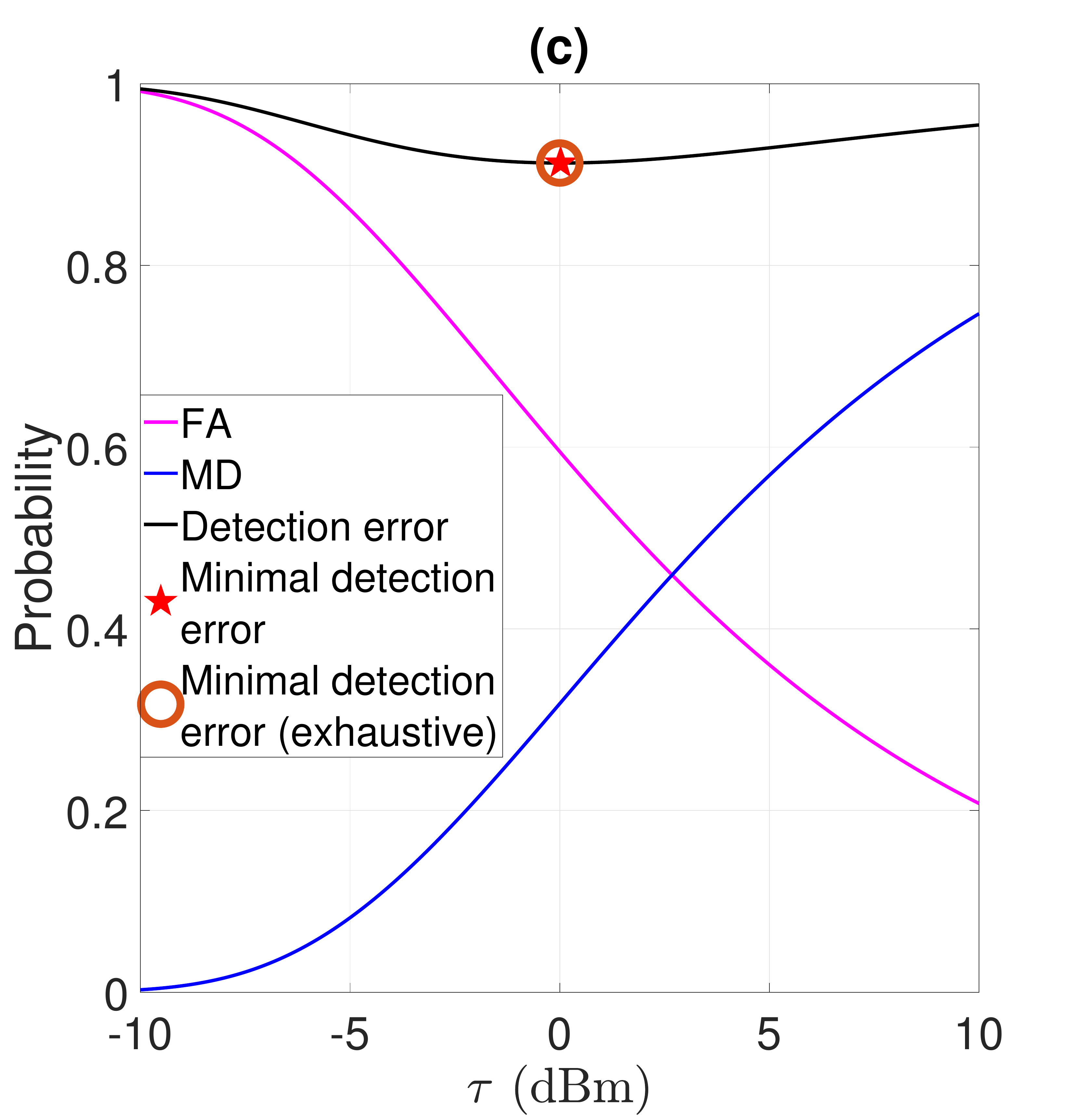}
     \end{minipage}
        \caption{Comparison between the simulation and analytic results on detection error with (a) $p^{\rm{C}} = 20 {\rm{dBm}}$ and (b) $p^{\rm{D}}= 15 {\rm{dBm}}$, and (c) optimality of the solution to the adversary's problem with $p^{\rm{C}} = 20 {\rm{dBm}}$ and $p^{\rm{D}}= 15 {\rm{dBm}}$.}
        \label{fig:simulation_vs_analysis_detection_error_uniqueness_sol_adversary}
\end{figure}

Fig.~\ref{fig:simulation_vs_analysis_detection_error_uniqueness_sol_adversary}(c) illustrates the FA probability, MD probability, and the detection error with adversary $a$'s detection threshold $\tau$ as the argument. We can observe that given the D2D and cellular transmission powers, e.g., $p^{\rm{D}} = 15 {\rm{dBm}}$ and $p^{\rm{C}} = 20 {\rm{dBm}}$ in Fig.~\ref{fig:simulation_vs_analysis_detection_error_uniqueness_sol_adversary}(c), the FA probability monotonically decreases while the MD probability monotonically increases as the detection threshold $\tau$ increases. This leads to a valley shape for the detection error with respect to (w.r.t.) the detection threshold $\tau$, which results in a global minimal detection error marked by the red pentagram in Fig.~\ref{fig:simulation_vs_analysis_detection_error_uniqueness_sol_adversary}(c). Such a minimum can also be observed when given $p^{\rm{C}}$ and varying $p^{\rm{D}}$ as shown in Fig.~\ref{fig:simulation_vs_analysis_detection_error_uniqueness_sol_adversary}(a) and vice versa in Fig.~\ref{fig:simulation_vs_analysis_detection_error_uniqueness_sol_adversary}(b). In this case, the existence and uniqueness of the global minimal detection error of adversary $a$ are verified. Moreover, due to the continuity and valley shape of the detection error w.r.t. the detection threshold $\tau$, numerous optimization methods are applicable to minimize it and obtain the global minimal detection error, and the corresponding detection threshold is the global optimal detection threshold $\tau^\star$. This global optimal detection threshold is the optimal strategy of adversary $a$ and thereby regarded as the best response from the lower stage.

\subsection{Upper-stage Legitimate Utility Maximization Problem Formulation and Solution Analysis}

\subsubsection{Problem Formulation}

At the upper stage, the legitimate entity aims to maximize its utility constrained by the D2D communication covertness and the cellular QoS as follows:
\begin{subequations}\label{eq:upper_problem}
\begin{align}
\max_{p^{\rm{D}}, \, p^{\rm{C}}} & \, \phi^{\rm{D}}{\mathbb{P}}\left[{\text{SINR}}_{d^{\rm{Tx}}} > \theta^{\rm{D}}\right] - \phi^{\rm{C}}\frac{\lambda_{\cal{U}}}{\lambda_{{\cal{D}}^{\rm{Tx}}}} {\mathbb{E}}\left[I^{\rm{D}}_u\right] \label{eq:obj}\\
{\text{s.\,t.}} & \, P^{\rm{FA}}_a\left(p^{\rm{D}}, p^{\rm{C}}, \tau^\star\right) + P^{\rm{MD}}_a\left(p^{\rm{D}}, p^{\rm{C}}, \tau^\star\right) \ge 1 - \varepsilon^{\rm{D}}\label{eq:constr_comm_covert}\\
& \, {\overline{R}}_u\left(p^{\rm{D}}, p^{\rm{C}}\right) \slash {\overline{\widetilde{R}}}_u\left(p^{\rm{D}}, p^{\rm{C}}\right) \ge 1 - \varepsilon^{\rm{C}}\label{eq:constr_cellular_qos}\\
& \, {\underline{p}}^{\rm{D}} \le p^{\rm{D}} \le {\overline{p}}^{\rm{D}}, \, {\underline{p}}^{\rm{C}} \le p^{\rm{C}} \le {\overline{p}}^{\rm{C}}.
\end{align}
\end{subequations}
Therein, the objective~(\ref{eq:obj}) is the legitimate utility and defined as the difference between the reward, i.e., $\phi^{\rm{D}}$, achieved from guaranteeing the D2D link reliability, i.e., ${\text{SINR}}_{d^{\rm{Tx}}} > \theta^{\rm{D}}$, and the cost incurred by the interference from the D2D network to the cellular network, i.e., $\phi^{\rm{C}}\frac{\lambda_{\cal{U}}}{\lambda_{{\cal{D}}^{\rm{Tx}}}} {\mathbb{E}}\left[I^{\rm{D}}_u\right]$, due to the underlaying scenario~\cite{7736071, 6554552}. Here, we set the cost proportional to the interference received by the CUs. In particular, let $A$ be the area of the region of interest, the number of CUs in the region is accordingly $\lambda_{\cal{U}} A$ and the total interference received by the CUs is $\lambda_{\cal{U}} A {\mathbb{E}}\left[I^{\rm{D}}_u\right]$, where ${\mathbb{E}}\left[I^{\rm{D}}_u\right]$ is the expectation of the interference from the D2D network to CU $u$ defined in~(\ref{eq:expectation_I_D_u}). We further let $\phi^{\rm{C}}$ be the price of causing one watt of interference to a CU and the total interference cost is therefore $\phi^{\rm{C}}\lambda_{\cal{U}} A {\mathbb{E}}\left[I^{\rm{D}}_u\right]$, which will be equally shared by all the D2D users in the region, i.e., $\lambda_{{\cal{D}}^{\rm{Tx}}} A$. In this case, each D2D user needs to pay $\phi^{\rm{C}}\frac{\lambda_{\cal{U}}A}{\lambda_{{\cal{D}}^{\rm{Tx}}}A} {\mathbb{E}}\left[I^{\rm{D}}_u\right] = \phi^{\rm{C}}\frac{\lambda_{\cal{U}}}{\lambda_{{\cal{D}}^{\rm{Tx}}}} {\mathbb{E}}\left[I^{\rm{D}}_u\right]$. The first inequality constraint~(\ref{eq:constr_comm_covert}) is to lower bound the detection error for adversary $a$ so as to ensure a certain D2D communication covertness, i.e., $1 - \varepsilon^{\rm{D}}$, where $\tau^\star$ is the best response from the lower stage corresponding to $p^{\rm{D}}$ and $p^{\rm{C}}$ and $\varepsilon^{\rm{D}}$ is the tolerable threshold of detection error. The second inequality constraint~(\ref{eq:constr_cellular_qos}) ensures that the degradation in the cellular ergodic rate caused by the interference from the underlaying D2D network is smaller than $\varepsilon^{\rm{C}}$ of the original cellular ergodic rate such that the cellular QoS will not be severely affected~\cite{5475350}. ${\underline{p}}^{\rm{D}}$ (${\underline{p}}^{\rm{C}}$) and ${\overline{p}}^{\rm{D}}$ (${\overline{p}}^{\rm{C}}$) are the lower and upper bounds of D2D (cellular) transmission power, respectively.

\subsubsection{Solution Analysis} 

Though the objective and constraints of the problem in~(\ref{eq:upper_problem}) have exact expressions and their values can be calculated via numerical integration techniques, it is too unwieldy to verify the convexity of the problem due to the multiple integrals in both the objective and constraints. Hence, we design a bi-level algorithm as shown in Algorithm~\ref{algo:bi-level} to solve the problem in~(\ref{eq:upper_problem}) based on SCA method and taking into account the best response from the lower stage. The solution obtained by the bi-level algorithm, i.e., the optimal strategy of the legitimate entity, together with the corresponding best response from the lower stage, i.e., the optimal strategy of the adversary, constitute the equilibrium of the two-stage Stackelberg game. In particular, let 
\begin{equation}
\begin{aligned}
\Xi_0 \left(p^{\rm{D}}, p^{\rm{C}}\right) = & -\phi^{\rm{D}}{\mathbb{P}}\left[{\text{SINR}}_{d^{\rm{Tx}}} > \theta^{\rm{D}}\right] + \phi^{\rm{C}}\frac{\lambda_{\cal{U}}}{\lambda_{{\cal{D}}^{\rm{Tx}}}} {\mathbb{E}}\left[I^{\rm{D}}_u\right],\\
\Xi_1 \left(p^{\rm{D}}, p^{\rm{C}}\right) = & -P^{\rm{FA}}_a\left(p^{\rm{D}}, p^{\rm{C}}, \tau^\star\right) - P^{\rm{MD}}_a\left(p^{\rm{D}}, p^{\rm{C}}, \tau^\star\right) + \left(1 - \varepsilon^{\rm{D}}\right), \\
\Xi_2 \left(p^{\rm{D}}, p^{\rm{C}}\right) = & -{\overline{R}}_u\left(p^{\rm{D}}, p^{\rm{C}}\right) + \left(1 - \varepsilon^{\rm{C}}\right) {\overline{\widetilde{R}}}_u\left(p^{\rm{D}}, p^{\rm{C}}\right),
\end{aligned}
\end{equation}
the expression of the problem given in~(\ref{eq:upper_problem}) is simplified as follows:
\begin{equation}\label{eq:simplified_upper_problem}
\begin{aligned}
\min_{p^{\rm{D}}, \, p^{\rm{C}}} & \, \Xi_0 \left(p^{\rm{D}}, p^{\rm{C}}\right)\\
{\text{s. t.}} & \, \Xi_i \left(p^{\rm{D}}, p^{\rm{C}}\right) \le 0, \, \forall i \in \left\{1, 2\right\}\\
& \, {\underline{p}}^{\rm{D}} \le p^{\rm{D}} \le {\overline{p}}^{\rm{D}}, \, {\underline{p}}^{\rm{C}} \le p^{\rm{C}} \le {\overline{p}}^{\rm{C}}.
\end{aligned}
\end{equation}
Here, we follow Algorithm~\ref{algo:bi-level} (see Section~2.1 of~\cite{razaviyayn2014successive} for detailed discussion) and switch to repetitively solve a local tight approximation of the problem in~(\ref{eq:simplified_upper_problem}) as follows:
\begin{equation}\label{eq:simplified_upper_problem_convex_approximation}
\begin{aligned}
\min_{p^{\rm{D}}, \, p^{\rm{C}}} & \, {\widehat{\Xi}}_0 \left(\left.p^{\rm{D}}, p^{\rm{C}}\right|p^{{\rm{D}}, \langle k-1 \rangle}, p^{{\rm{C}}, \langle k-1 \rangle}\right)\\
{\text{s. t.}} & \, {\widehat{\Xi}}_i \left(\left.p^{\rm{D}}, p^{\rm{C}}\right|p^{{\rm{D}}, \langle k-1 \rangle}, p^{{\rm{C}}, \langle k-1 \rangle}\right) \le 0, \, \forall i \in \left\{1, 2\right\}\\
& \, {\underline{p}}^{\rm{D}} \le p^{\rm{D}} \le {\overline{p}}^{\rm{D}}, \, {\underline{p}}^{\rm{C}} \le p^{\rm{C}} \le {\overline{p}}^{\rm{C}},
\end{aligned}
\end{equation}
where $\left[p^{{\rm{D}}, \langle k-1 \rangle}, p^{{\rm{C}}, \langle k-1 \rangle}\right]$ is the solution obtained in iteration $k - 1$. Therein,
\begin{equation}
\begin{aligned}
& {\widehat{\Xi}}_i \left(\left.p^{\rm{D}}, p^{\rm{C}}\right|p^{{\rm{D}}, \langle k-1 \rangle}, p^{{\rm{C}}, \langle k-1 \rangle}\right) \\
\triangleq & \Xi_i \left(p^{{\rm{D}}, \langle k-1 \rangle}, p^{{\rm{C}}, \langle k-1 \rangle}\right) + \left.\nabla \Xi_i \left(p^{{\rm{D}}}, p^{{\rm{C}}}\right)\right|_{p^{\rm{D}} = p^{{\rm{D}}, \langle k-1 \rangle}, p^{\rm{C}} = p^{{\rm{C}}, \langle k-1 \rangle}}
\left(
\begin{aligned}
p^{\rm{D}} - p^{{\rm{D}}, \langle k-1 \rangle}\\
p^{\rm{C}} - p^{{\rm{C}}, \langle k-1 \rangle}
\end{aligned}
\right) \\
& + \frac{1}{2\delta} \left(
\begin{aligned}
p^{\rm{D}} - p^{{\rm{D}}, \langle k-1 \rangle}\\
p^{\rm{C}} - p^{{\rm{C}}, \langle k-1 \rangle}
\end{aligned}
\right)^\top \left(
\begin{aligned}
p^{\rm{D}} - p^{{\rm{D}}, \langle k-1 \rangle}\\
p^{\rm{C}} - p^{{\rm{C}}, \langle k-1 \rangle}
\end{aligned}
\right)
\end{aligned}
\end{equation}
for all $i\in \left\{0, 1, 2\right\}$ follows the gradient descent method, i.e., ${\widehat{\Xi}}_i \left(\left.p^{\rm{D}}, p^{\rm{C}}\right|p^{{\rm{D}}, \langle k-1 \rangle}, p^{{\rm{C}}, \langle k-1 \rangle}\right)$ is consistent with $\Xi_i \left(p^{\rm{D}}, p^{\rm{C}}\right)$ in both function value and gradient over $\left[p^{{\rm{D}}, \langle k-1 \rangle}, p^{{\rm{C}}, \langle k-1 \rangle}\right]$, where $\delta$ is small enough such that ${\widehat{\Xi}}_i \left(\left.p^{\rm{D}}, p^{\rm{C}}\right|p^{{\rm{D}}, \langle k-1 \rangle}, p^{{\rm{C}}, \langle k-1 \rangle}\right) \ge \Xi_i \left(p^{\rm{D}}, p^{\rm{C}}\right)$.

\begin{algorithm}[!ht]
\DontPrintSemicolon
  \KwData{Initialize $p^{{\rm{D}}, \langle k \rangle}$ and $p^{{\rm{C}}, \langle k \rangle}$ with $k = 0$ and set the convergence tolerance $\epsilon$.}
   \While{not convergent}
   {Solve problem in~(\ref{eq:simplified_upper_problem_convex_approximation}) and obtain the temporary solution $p^{{\rm{D}}, {\text{temp}}}$ and $p^{{\rm{C}}, {\text{temp}}}$ at iteration $k$;\;
   Update the solution with $p^{{\rm{D}}, {\langle k + 1 \rangle}} = p^{{\rm{D}}, {\langle k \rangle}} \gamma + p^{{\rm{D}}, {\text{temp}}} \left(1 - \gamma\right)$ and $p^{{\rm{C}}, {\langle k + 1 \rangle}} = p^{{\rm{C}}, {\langle k \rangle}} \gamma + p^{{\rm{C}}, {\text{temp}}} \left(1 - \gamma\right)$;\;
   Update $k$ by $k + 1 \mapsto k$;\;
   }
\caption{Bi-Level Algorithm}
\label{algo:bi-level}
\end{algorithm}


\section{Performance Evaluation}
\label{sec:performance}

In this section, we present the experimental results to evaluate the performance of the covert communication in the D2D-underlaid cellular network. The parameters are given in Table~\ref{tab:notation_value}. First, we demonstrate the assistance from the cellular network in achieving covert communication on the D2D transmission in Fig.~\ref{fig:role_cellular}. Second, we evaluate the impact of the constraints on D2D communication covertness and cellular QoS on the system performance in Fig.~\ref{fig:impact_varepsilon_D_C}. Third, the impact of the D2D communication distance and adversaries' density on the system performance has been evaluated in Fig.~\ref{fig:impact_R_lambda_A}. Fourth, we illustrate the impact of D2D transmitters' density and BSs' density on the legitimate utility. Fifth, we show and analyze the impact of the SINR threshold of the D2D communication and D2D transmitters' density on the legitimate utility.

\begin{table*}[!]
	\centering
	\caption{Symbols and descriptions}
	\begin{tabular}{|c|l|c|}
		\hline
		\hline
		{\bf{Symbol}} & {\bf{Description}} & {\bf{Default value}}\\
		\hline
		\makecell[c]{$d^{\rm{Tx}}$, $d^{\rm{Rx}}$, \\$u$, $b$, $a$} & \makecell[l]{Indexes of typical D2D transmitter, typical D2D receiver, typical \\CU, typical BS, and typical adversary, respectively.} & --- \\
		\hline
		\makecell[c]{${\bf{x}}_{d^{\rm{Tx}}}$, ${\bf{x}}_{d^{\rm{Rx}}}$, \\${\bf{x}}_{u}$, ${\bf{x}}_{b}$, ${\bf{x}}_{a}$} & \makecell[l]{Locations of typical D2D transmitter, typical D2D receiver, \\typical CU, typical BS, and typical adversary, respectively.} & --- \\
        \hline
		${\cal{D}}^{\rm{Tx}}$, ${\cal{B}}$, ${\cal{U}}$, ${\cal{A}}$ & Sets of D2D transmitters, BSs, CUs, and adversaries, respectively. & --- \\
        \hline
		\makecell[c]{$\Phi_{{\cal{D}}^{\rm{Tx}}}$, $\Phi_{\cal{B}}$, \\$\Phi_{\cal{U}}$, $\Phi_{\cal{A}}$} & PPs of D2D transmitters, BSs, CUs, and adversaries, respectively. & --- \\
		\hline
		\makecell[c]{$\lambda_{{\cal{D}}^{\rm{Tx}}}$, $\lambda_{\cal{B}}$, \\$\lambda_{\cal{U}}$, $\lambda_{\cal{A}}$} & Densities of $\Phi_{{\cal{D}}^{\rm{Tx}}}$, $\Phi_{\cal{B}}$, $\Phi_{\cal{U}}$, and $\Phi_{\cal{A}}$, respectively. & \makecell[c]{$0.1$, $0.01$, \\$0.1$, $0.01$~\cite{7893755}.}\\	
		\hline
		${\cal{D}}_1$, ${\cal{C}}_1$ & Events that D2D transmitter and BS are active, respectively. & --- \\	
        \hline
		${\mathbb{P}}^{{\cal{D}}_1}$, ${\mathbb{P}}^{{\cal{C}}_1}$ & \makecell[l]{Probabilities of events ${\cal{D}}_1$ and ${\cal{C}}_1$, respectively.} & $0.3$, $0.7$~\cite{7893755}. \\	
		\hline
		$p^{\rm{D}}$, $p^{\rm{C}}$, $\tau$ & \makecell[l]{D2D transmission power, cellular transmission power, and power \\detection threshold, respectively.} & \makecell[c]{$\left[0, 20\right]\,{\rm{dBm}}$~\cite{9169911}, \\$\left[-10, 10\right]\,{\rm{dBW}}$, ---.} \\
		\hline
		$h_{{d^{\rm{Tx}}} u}$, $R$ & \makecell[l]{Small-scale fading gain between D2D transmitter ${d^{\rm{Tx}}}$ and CU $u$ \\and link distance between D2D transmitter and its dedicated \\D2D receiver, respectively.} & $\exp\left(1\right)$, $1$.\\
		\hline
		\makecell[c]{$\alpha$, $\varepsilon^{\rm{D}}$, $\varepsilon^{\rm{C}}$, \\$\theta^{\rm{D}}$, $\theta^{\rm{C}}$} & \makecell[l]{Path-loss exponent, threshold of detection error, ratio threshold \\of rate loss at CU, SINR threshold at D2D receiver, and SINR \\threshold at CU, respectively.} & \makecell[c]{$4$, $0.01$, $0.05$,\\ $-10\,{\rm{dB}}$, ---.} \\
		\hline
		$N_{d^{\rm{Rx}}}$, $N_u$, $N_a$ & \makecell[l]{Additive noise at D2D receiver $d^{\rm{Rx}}$, CU $u$, and adversary $a$, \\respectively.} & $- 90\,{\rm{dBm}}$. \\
		\hline
		$\phi^{\rm{D}}$, $\phi^{\rm{C}}$ & \makecell[l]{Reward of guaranteeing the D2D link reliability and cost of \\incurring interference to cellular communication, respectively.} & $10$, $0.05$.\\
		\hline
		$f_y$, $F_y$ & PDF and CDF of a random variable $y$, respectively. & --- \\
		\hline
	\end{tabular}
	\label{tab:notation_value}
\end{table*}


\subsection{Role of Cellular Network}

\begin{figure}
     \centering
     \begin{minipage}{8cm}
		\centering
		\includegraphics[width=1\textwidth,trim=10 0 70 20,clip]{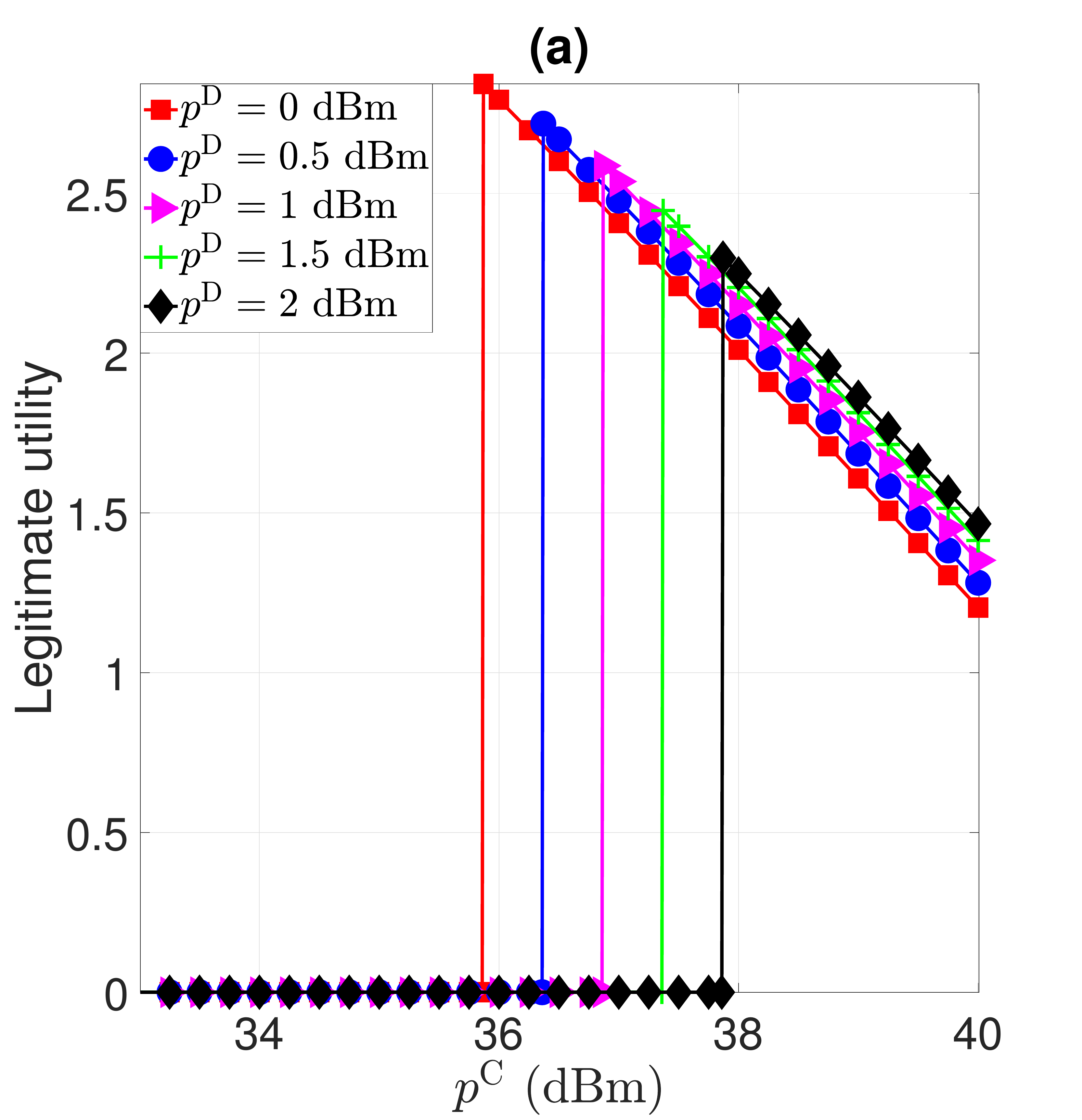}
     \end{minipage}
     \begin{minipage}{8cm}
		\centering
		\includegraphics[width=1\textwidth,trim=10 0 70 20,clip]{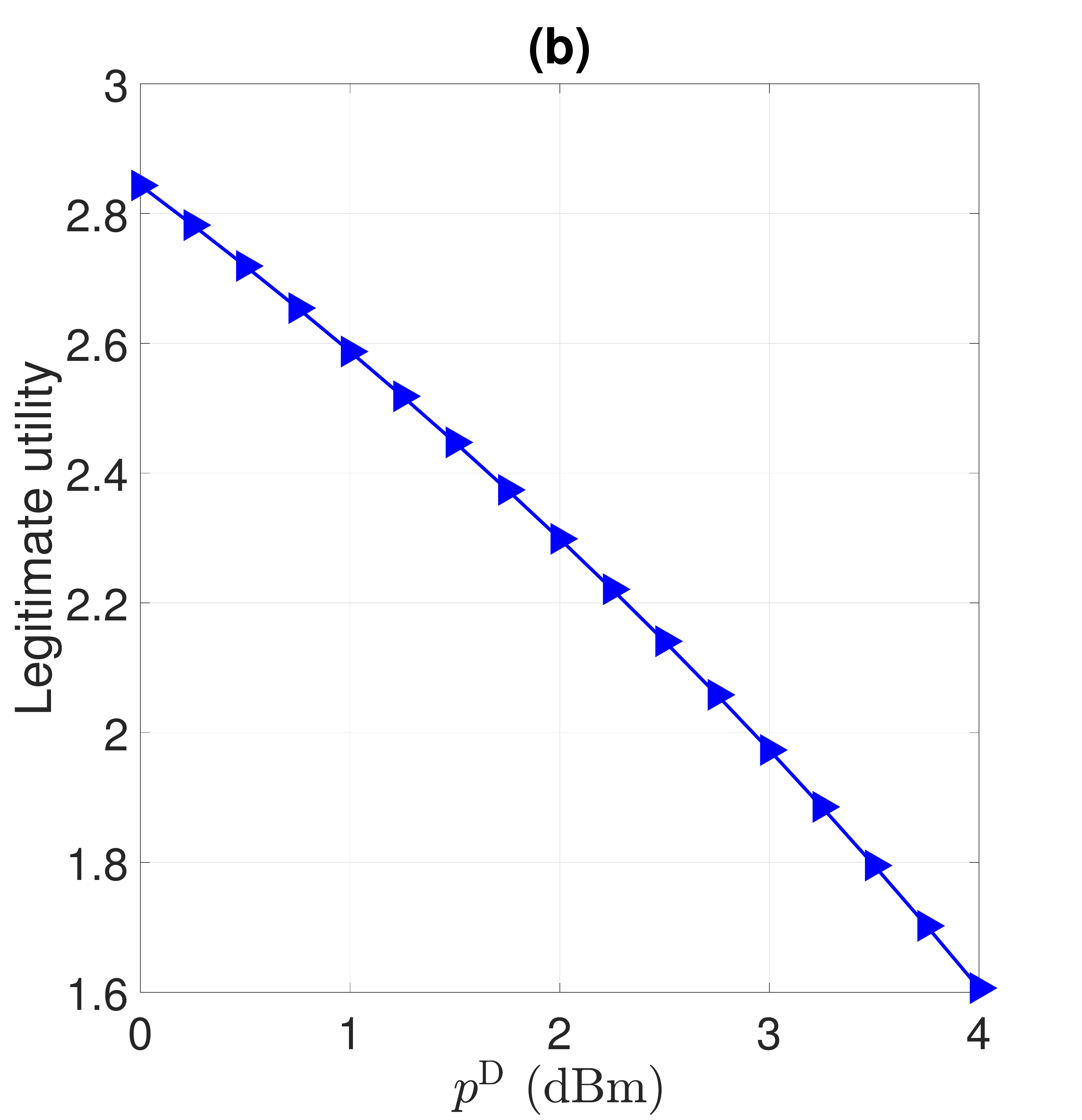}
     \end{minipage}
        \caption{(a) Legitimate utility versus cellular transmission power $p^{{\rm{C}}}$ with varying D2D transmission power $p^{{\rm{D}}}$ and (b) legitimate utility versus D2D transmission power $p^{{\rm{D}}}$.}
        \label{fig:role_cellular}
\end{figure}

Figure~\ref{fig:role_cellular} demonstrates the role of the cellular network in achieving covert D2D transmission. In Fig.~\ref{fig:role_cellular}(a), we can observe that the legitimate utility is $0$ when the cellular transmission power $p^{{\rm{C}}}$ is small, e.g., $p^{{\rm{C}}} \lessapprox 37 {\rm{dBm}}$\footnote{$\lessapprox$ means being approximately smaller than.} when D2D transmission power $p^{{\rm{D}}} = 1 {\rm{dBm}}$. This is due to the fact that when $p^{{\rm{C}}}$ is small, the interference from the cellular network is too weak to distort the observation and further mislead the decision of the adversary and hence the constraint on D2D communication covertness cannot be met. In this case, it is unable to achieve covert communication on the D2D transmission and hence the legitimate utility is $0$. Later, when $p^{{\rm{C}}}$ approaches a certain level, e.g., $p^{{\rm{C}}} \gtrapprox 37 {\rm{dBm}}$\footnote{$\gtrapprox$ means being approximately larger than.} when D2D transmission power $p^{{\rm{D}}} = 1 {\rm{dBm}}$, the legitimate utility becomes nonzero and decreases along with the increase in $p^{{\rm{C}}}$. The reason for the decrease in the nonzero legitimate utility is due to the increasing interference from the cellular network to the D2D communication. In this case, there is a trade-off in leveraging the assistance from the cellular network to achieve covert communication on D2D transmission. Additionally, as shown in Fig.~\ref{fig:role_cellular}(b), the increase in $p^{{\rm{D}}}$ damages the legitimate utility. The reason can be explained as follows. To achieve covert communication on the D2D transmission when $p^{{\rm{D}}}$ increases, we have to increase $p^{{\rm{C}}}$. Moreover, the increasing speed in $p^{{\rm{C}}}$ should be faster than that in $p^{{\rm{D}}}$ so as to maintain a certain level of covertness for the D2D communication. In this case, the interference from the cellular network to the D2D communication increases faster than $p^{{\rm{D}}}$, which induces a weaker SINR for the D2D transmission and further a lower legitimate utility as shown in Fig.~\ref{fig:role_cellular}(b).


\subsection{Impact of the Constraint on D2D Communication Covertness and Cellular QoS}

\begin{figure}[!]
	\centering
	\includegraphics[width=0.5\textwidth,trim=10 10 40 60, clip]{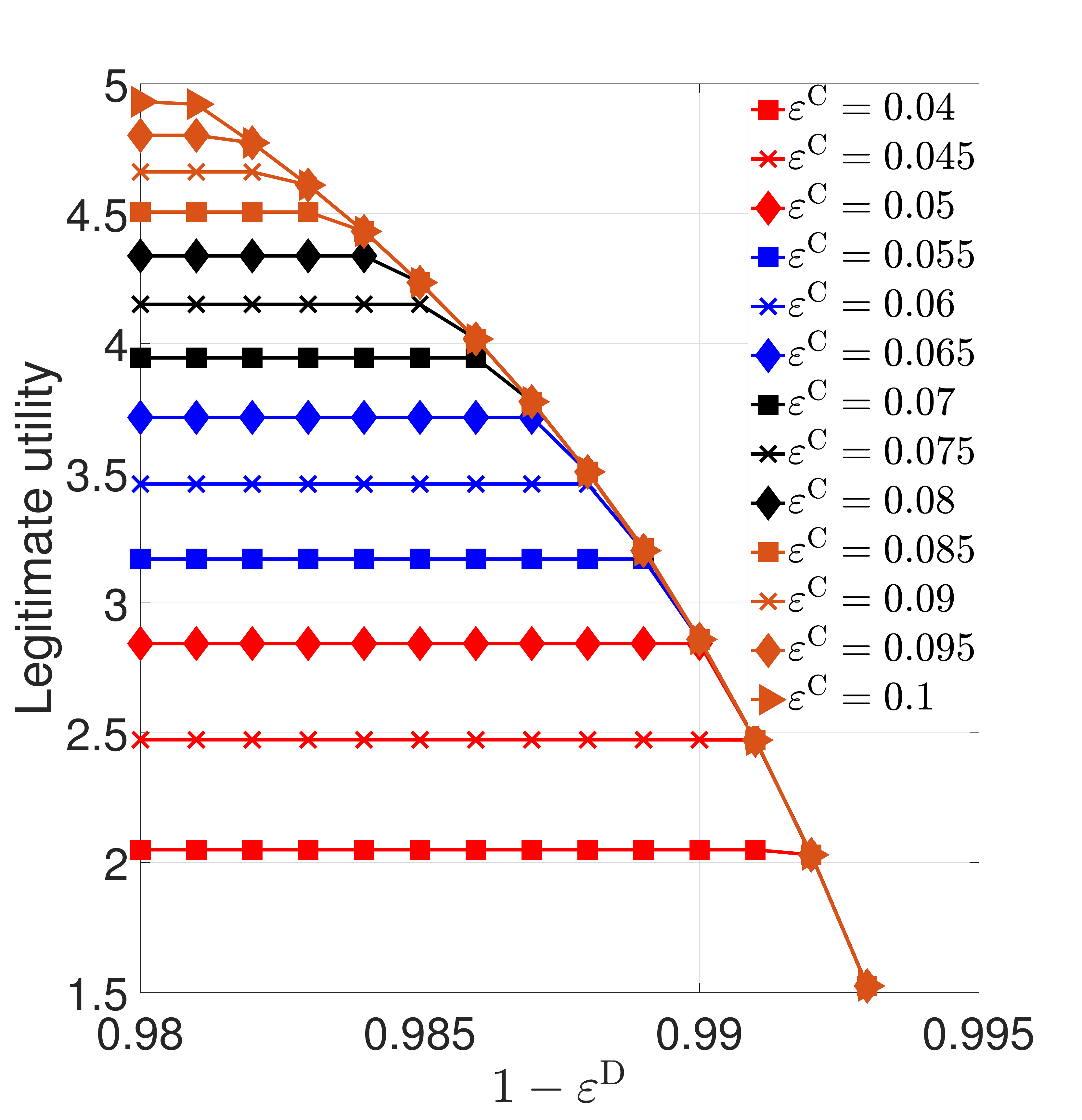}
	\caption{Legitimate utility versus the constraint on D2D communication covertness $1 - \varepsilon^{\rm{D}}$ with different constraints on cellular QoS $\varepsilon^{\rm{C}}$.}
	\label{fig:impact_varepsilon_D_C}
\end{figure}

Figure~\ref{fig:impact_varepsilon_D_C} shows the impact of the constraint on D2D communication covertness $1 - \varepsilon^{\rm{D}}$ and that on cellular QoS $\varepsilon^{\rm{C}}$ on the legitimate utility. Initially, as the constraint on D2D communication covertness becomes tighter, the legitimate utility remains the same, e.g., legitimate utility stays around $2.5$ when $1 - \varepsilon^{\rm{D}}$ increases from $0.98$ to $0.991$ subject to the constraint on cellular QoS $\varepsilon^{\rm{C}} = 0.045$. The reason is that the constraint on cellular QoS also limits the D2D transmission power $p^{{\rm{D}}}$ and is tighter than that on D2D communication covertness when $\varepsilon^{\rm{C}} = 0.045$ and $1 - \varepsilon^{\rm{D}} \in \left[0.98,0.991\right]$. In this case, the constraint on D2D communication covertness does not take effect and hence the legitimate utility remains the same. Later, when $1 - \varepsilon^{\rm{D}}$ further increases, the constraint on D2D communication covertness becomes tighter than that on cellular QoS and starts to take effect, e.g., the legitimate utility decreases when $\varepsilon^{\rm{C}} = 0.045$ and $1 - \varepsilon^{\rm{D}} \in \left[0.991,0.993\right]$. The insight revealed in the above discussion is that the constraint on D2D communication covertness, i.e.,~(\ref{eq:constr_comm_covert}), has a similar impact on the D2D transmission power $p^{\rm{D}}$ as that on cellular QoS, i.e.,~(\ref{eq:constr_cellular_qos}). In this case, the power control on the D2D transmission power $p^{\rm{D}}$ is subject to either of the aforementioned constraints, i.e.,~(\ref{eq:constr_comm_covert}) or~(\ref{eq:constr_cellular_qos}), depending on whichever is tighter.


\subsection{Impact of D2D communication distance and adversaries' density}

\begin{figure}[!]
	\centering
	\includegraphics[width=0.5\textwidth,trim=10 10 60 60, clip]{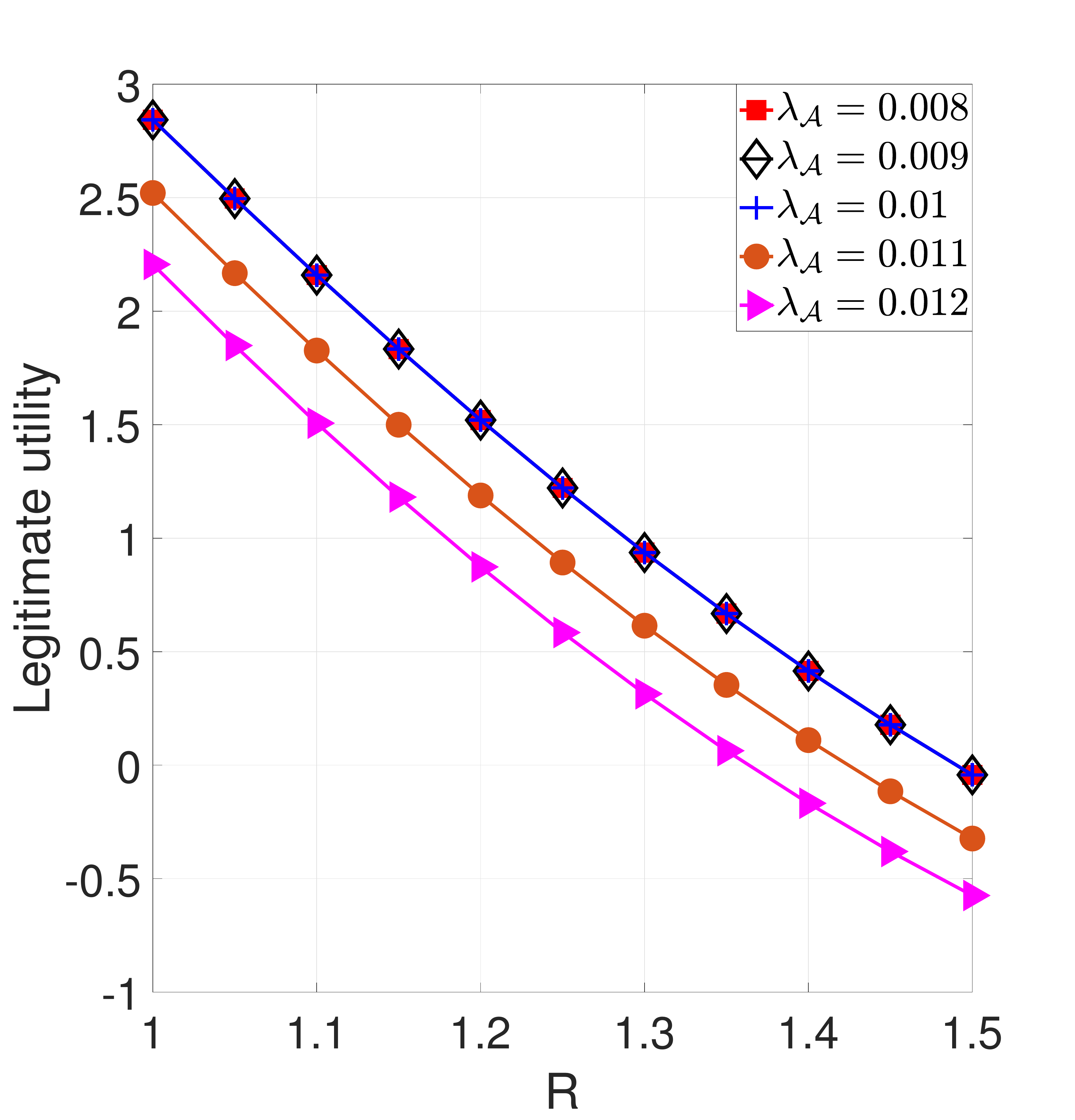}
	\caption{Legitimate utility versus the D2D communication distance $R$ with different adversaries' densities $\lambda_{\cal{A}}$.}
	\label{fig:impact_R_lambda_A}
\end{figure}

Figure~\ref{fig:impact_R_lambda_A} shows the impact of D2D communication distance $R$ and adversaries' density $\lambda_{\cal{A}}$. We can observe that the legitimate utility decreases as the D2D communication distance $R$ increases. This is due to due to several reasons. First, the link reliability decreases as the D2D communication distance increases. Second, with a larger D2D communication distance, if the D2D transmission power is increased to maintain the link reliability, it is easier for the adversary to detect the D2D transmission, which will also damage the legitimate utility. This additionally and adversely implied that decreasing D2D communication distance improves not only the link reliability but also the D2D communication covertness. Hence, in the networks with cooperative relays, higher densification of relay nodes will significantly improve the D2D communication covertness and achieve better network security. In addition, an interesting result is that the legitimate utility remains the same when the adversaries' density $\lambda_{\cal{A}}$ decreases from $0.01$ to $0.008$. The reason can be explained as follows. When the adversaries' density $\lambda_{\cal{A}}$ decreases from $0.01$ to $0.008$, the distance between the adversary and the D2D transmitter becomes larger, which increases difficulty in the transmission detection for the adversary. In this case, intuitively, the constraint on D2D communication covertness can still be satisfied even when the D2D transmission power $p^{{\rm{D}}}$ is increased, which will induce a higher legitimate utility. However, due to the constraint on cellular QoS, the D2D transmission power cannot be increased to a higher level, which results in an unchanged legitimate utility as shown in Fig.~\ref{fig:impact_R_lambda_A}.


\subsection{Impact of BSs' Density and D2D Transmitters' Density}

\begin{figure}[!]
	\centering
	\includegraphics[width=0.5\textwidth,trim=10 10 40 60, clip]{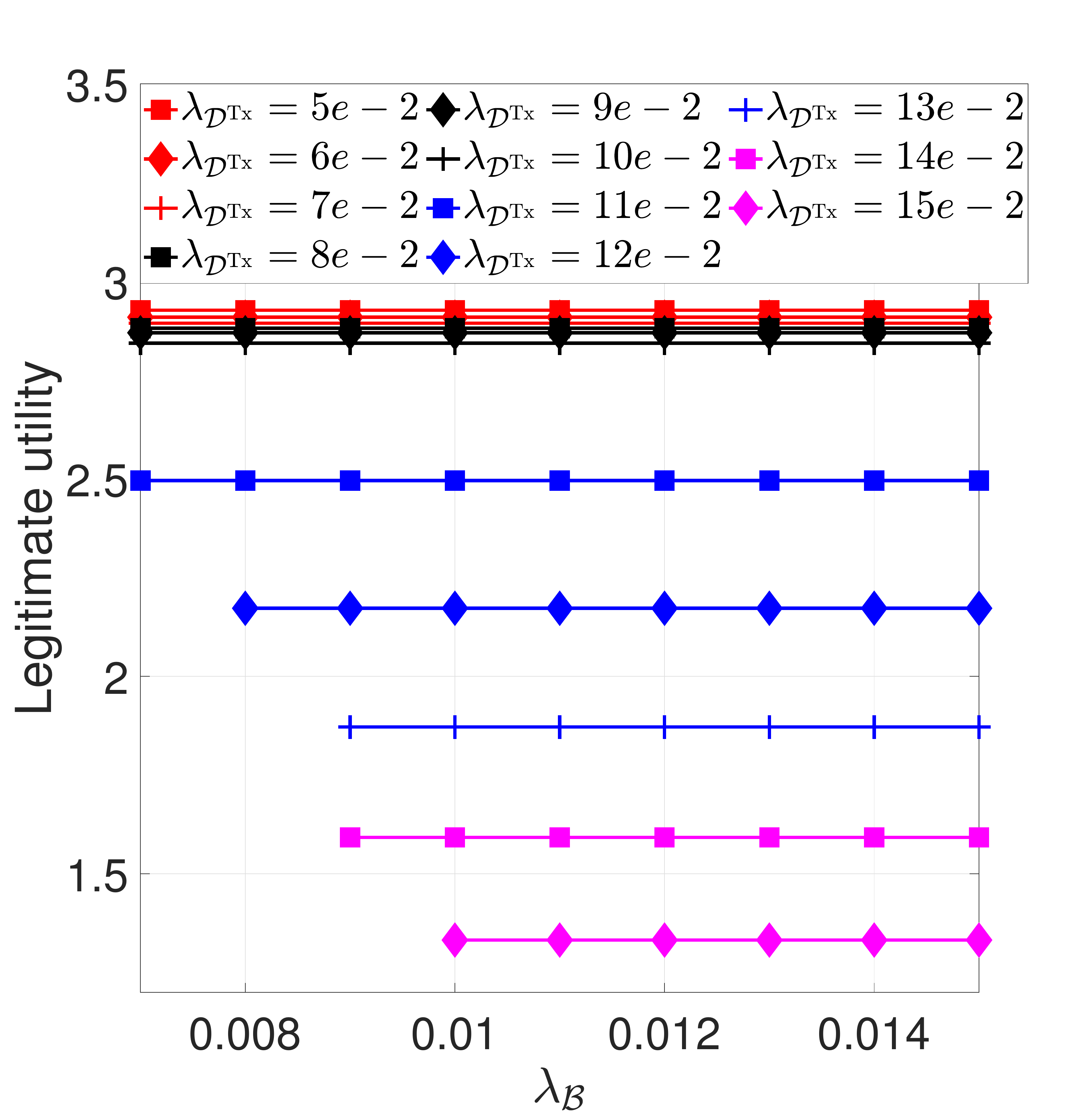}
	\caption{Legitimate utility versus the BSs' density $\lambda_{\cal{B}}$ with different D2D transmitters' densities $\lambda_{{\cal{D}}^{\rm{Tx}}}$.}
\label{fig:impact_lambda_B_DTX}
\end{figure}

We evaluate the impact of BSs' density $\lambda_{\cal{B}}$ and D2D transmitters' density $\lambda_{{\cal{D}}^{\rm{Tx}}}$ in Fig~\ref{fig:impact_lambda_B_DTX}. When the D2D transmitters' density $\lambda_{{\cal{D}}^{\rm{Tx}}} = 0.13$, the legitimate utility does not exist until the BSs' density $\lambda_{\cal{B}} \ge 0.009$. This is due to the fact that the interference from the cellular network is insufficient to distort the adversary's observation when the BSs' density is too small, e.g., $\lambda_{\cal{B}} < 0.009$ when $\lambda_{{\cal{D}}^{\rm{Tx}}} = 0.13$. In this case, the constraint on D2D communication covertness cannot be satisfied and the legitimate utility does not exist. This again validates the necessity of leveraging the cellular signal to improve the D2D communication covertness. In addition, the decrease in legitimate utility corresponding to the increase in D2D transmitters' density is non-uniform, i.e., legitimate utility decreases approximately less than $0.2$ when D2D transmitters' density $\lambda_{{\cal{D}}^{\rm{Tx}}}$ increases from $0.05$ to $0.1$ while decreases approximately $1.2$ when D2D transmitters' density $\lambda_{{\cal{D}}^{\rm{Tx}}}$ increases from $0.11$ to $0.15$. The reason can be explained as follows. First, when the D2D transmitters' density is small, i.e., $\left[0.05, 0.1\right]$, the interference from the D2D network to the cellular network is tiny, and hence the decrease in the legitimate utility corresponding to the increase in the D2D transmitters' density is small. Later, when D2D transmitters' density further increases, it will not only introduce much more interference to the cellular network and thereafter incur higher interference cost but also will require higher cellular transmission power to jointly ensure the constraints on D2D communication covertness and cellular QoS. This will correspondingly bring more interference to D2D communication. In this case, the legitimate utility reduces faster than ever. Additionally, the legitimate utility remains flat w.r.t. the BSs' density when the D2D transmitters' density is fixed. The reason can be explained as follows. Along with the increase in the BSs' density, the distance between the BS and its serving CU shortens and the cellular transmission power can be correspondingly reduced while the constraint on cellular QoS can still be satisfied. In this case, the interference from the cellular network remains the same, which results in an unchanged legitimate utility. 


\subsection{Impact of SINR thresholds at D2D receiver and D2D Transmitters' Density}

\begin{figure}[!]
	\centering
	\includegraphics[width=0.5\textwidth,trim=10 5 40 60, clip]{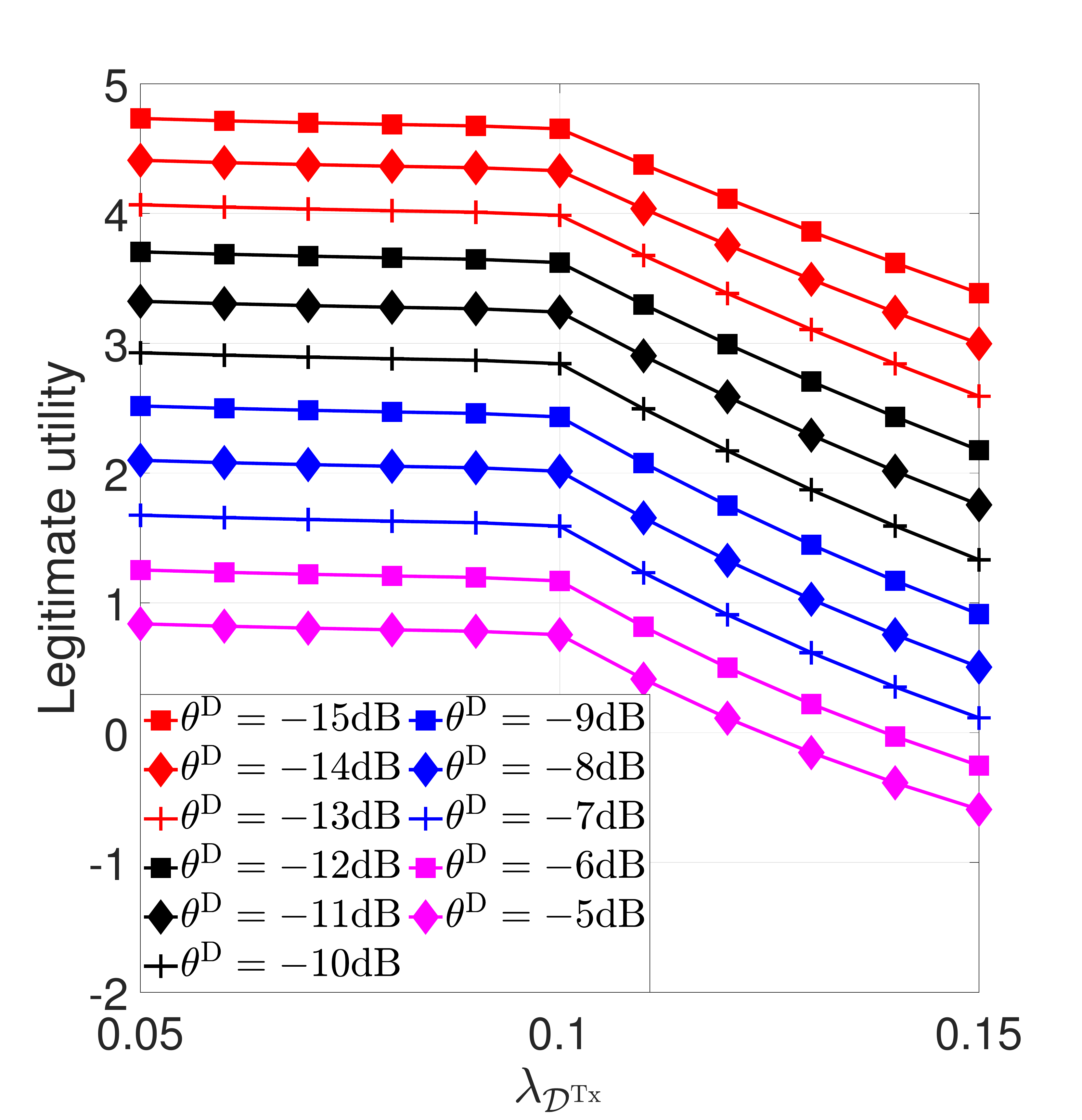}
	\caption{Legitimate utility versus the D2D transmitters' densities $\lambda_{{\cal{D}}^{\rm{Tx}}}$ with different SINR thresholds at D2D receiver $\theta^{{\rm{D}}}$.}
	\label{fig:impact_theta_D_lambda_DTx}
\end{figure}

Figure~\ref{fig:impact_theta_D_lambda_DTx} evaluates the impact of the D2D transmitters' densities $\lambda_{{\cal{D}}^{\rm{Tx}}}$ and the SINR thresholds at D2D receiver $\theta^{{\rm{D}}}$. We can observe that legitimate utility slowly decreases w.r.t. the D2D transmitters' densities $\lambda_{{\cal{D}}^{\rm{Tx}}}$ when $\lambda_{{\cal{D}}^{\rm{Tx}}} \le 0.1$ while fast decreases w.r.t. $\lambda_{{\cal{D}}^{\rm{Tx}}}$ when $\lambda_{{\cal{D}}^{\rm{Tx}}} > 0.1$. The reason can be explained as follows. When $\lambda_{{\cal{D}}^{\rm{Tx}}} \le 1$, both the distances between the adversaries and their target D2D transmitters and that between the CUs and the D2D transmitters are large. In this case, the constraint on D2D communication covertness is tighter than that on cellular QoS and hence takes effect. To meet such a constraint, we can slightly increase the cellular transmission power along with the increase in $\lambda_{{\cal{D}}^{\rm{Tx}}}$. Later, when $\lambda_{{\cal{D}}^{\rm{Tx}}} > 1$, as the distances between the D2D transmitters and CUs shorten and the interference from the D2D network to the cellular network dramatically increases, the constraint on cellular QoS is tighter than that on D2D communication covertness and starts to take effect. In this case, we have to dramatically increase the cellular transmission power, which significantly weakens the SINR of the D2D transmission and hence damages the legitimate utility. 


\section{Conclusion and Insights}
\label{sec:conclusion}

In this work, we have studied a D2D-underlaid cellular network and applied the covert communication to hide the presence of the D2D communication so as to defend against the adversary from transmission detection. We have modeled the combat between the adversary and the legitimate entity, i.e., D2D-underlaid cellular network, in the framework of two-stage Stackelberg game. Therein, the adversary is the follower at the lower stage and aims to minimize its detection error while the legitimate entity is the leader at the upper stage and aims to maximize its constrained utility. To conduct the study from the system-level perspective, we have modeled the spatial configuration of the D2D-underlaid cellular network and adversaries by stochastic geometry. To obtain the equilibrium of the proposed Stackelberg game, we first analyze the problem of the adversary and obtain its optimal strategy as the best response from the lower stage, and the optimality of which has been both numerically and theoretically verified. Then, taking into account the best response from the lower stage, we have designed a bi-level algorithm based on SCA method to search for the optimal strategy of the legitimate entity, which together with the optimal strategy of the adversary constitute the game equilibrium. Numerical results have been presented and interesting insights have been discussed, including
\begin{itemize}
\item There is a trade-off in leveraging the interference from the cellular network to achieve covert communication on the D2D transmission (see the discussion of Fig.~\ref{fig:role_cellular}). 

\item Decreasing the communication distance between the D2D users improves not only the link reliability but also the communication covertness of the D2D transmission (see the discussion of Fig.~\ref{fig:impact_R_lambda_A}).

\item The constraint on cellular QoS has similar impact on the legitimate utility as that on D2D communication covertness (see the discussion of Fig.~\ref{fig:impact_varepsilon_D_C}).
\end{itemize}
For future work, we will investigate the combat between the adversary and the legitimate entity on a long-run manner. 


\bibliography{bibfile}

\end{document}